\documentclass[fleqn,usenatbib]{mnras}

\usepackage{mathptmx}

\usepackage[T1]{fontenc}
\usepackage{ae,aecompl}

\usepackage{graphicx}	
\usepackage{amsmath}	
\usepackage{amssymb}	
\usepackage{color}
\usepackage[table]{xcolor}
\usepackage{multicol}

\newcommand{\app}[1]{Appendix~\ref{sec:#1}}

\newcommand{\fig}[1]{Figure~\ref{fig:#1}}
\renewcommand{\sec}[1]{Section~\ref{sec:#1}}
\newcommand{\tab}[1]{Table~\ref{tab:#1}}

\newcommand{\eagle}{\mbox{\sc{EAGLE}}}
\newcommand{\ceagle}{\mbox{\sc{C-EAGLE}}}
\newcommand{\flares}{\mbox{\sc{Flares}}}
\newcommand{\galform}{\mbox{\sc{Galform}}}
\newcommand{\pmill}{\mbox{\sc{P-Millennium}}}

\title[Mapping baryons onto dark matter haloes]{A machine learning approach to mapping baryons onto dark matter haloes using the \eagle\ and \ceagle\ simulations}

\author[Christopher C. Lovell et al.]{Christopher C. Lovell,$^{1}$\thanks{E-mail: c.lovell@herts.ac.uk (CCL)}
Stephen M. Wilkins,$^{2}$
Peter A. Thomas,$^{2}$
Matthieu Schaller,$^{3,4}$\newauthor 
Carlton M. Baugh,$^{5}$
Giulio Fabbian,$^{6,7}$ 
Yannick Bah\'{e}$^{4}$
\\
$^{1}$Centre for Astrophysics Research, School of Physics, Astronomy \& Mathematics, University of Hertfordshire, Hatfield AL10 9AB\\
$^{2}$Astronomy Centre, Department of Physics and Astronomy, University of Sussex, Brighton, BN1 9QH, UK\\
$^{3}$Lorentz Institute for Theoretical Physics, Leiden University, PO Box 9506, NL-2300 RA Leiden, The Netherlands\\
$^{4}$Leiden Observatory, Leiden University, PO Box 9513, NL-2300 RA Leiden, The Netherlands\\
$^{5}$Institute for Computational Cosmology, Department of Physics, Durham University, Science Laboratories, South Road, Durham, DH1 3LE, UK\\
$^{6}$School of Physics and Astronomy, Cardiff University, The Parade, Cardiff, CF24 3AA, UK\\
$^{7}$Center for Computational Astrophysics, Flatiron Institute, New York, 10010, NY, USA
}

\date{Accepted XXX. Received YYY; in original form ZZZ}

\pubyear{2021}

\begin{document}
\label{firstpage}
\pagerange{\pageref{firstpage}--\pageref{lastpage}}
\maketitle

\begin{abstract}
High-resolution cosmological hydrodynamic simulations are currently limited to relatively small volumes due to their computational expense.
However, much larger volumes are required to probe rare, overdense environments, and measure clustering statistics of the large scale structure.
Typically, zoom simulations of individual regions are used to study rare environments, and semi-analytic models and halo occupation models applied to dark matter only (DMO) simulations are used to study the Universe in the large-volume regime.
We propose a new approach, using a machine learning framework to explore the halo-galaxy relationship in the periodic \eagle\ simulations, and zoom \ceagle\ simulations of galaxy clusters.
We train a tree based machine learning method to predict the baryonic properties of galaxies based on their host dark matter halo properties.
The trained model successfully reproduces a number of key distribution functions for an infinitesimal fraction of the computational cost of a full hydrodynamic simulation.
By training on both periodic simulations as well as zooms of overdense environments, we learn the bias of galaxy evolution in differing environments.
This allows us to apply the trained model to a \textit{larger} DMO volume than would be possible if we only trained on a periodic simulation.
We demonstrate this application using the $(800 \; \mathrm{Mpc})^3$ P-Millennium simulation, and present predictions for key baryonic distribution functions and clustering statistics from the \eagle\ model in this large volume.
\end{abstract}




\section{Introduction}
\label{sec:intro}

Cosmological hydrodynamic simulations self-consistently model the evolution of baryonic and cold dark matter, and the subsequent hierarchical assembly of galaxies in a $\Lambda$CDM universe \citep{benson_galaxy_2010, somerville_physical_2015}.
A number of projects, such as \eagle\ \citep{schaye_eagle_2015}, \textsc{Illustris} \citep{vogelsberger_introducing_2014}, \textsc{Illustris-TNG} \citep{pillepich_simulating_2018}, \textsc{Mufasa} \citep{dave_mufasa:_2016} and \textsc{Simba} \citep{dave_simba:_2019} have had reasonable success at reproducing key galaxy distribution functions in the low-redshift Universe, such as the galaxy stellar mass function.
They are typically run within large periodic volumes, $\sim$100 Mpc on a side, and have mass resolutions of order $\sim 10^6 \, \mathrm{M_{\odot}}$.
This is sufficient to resolve the internal structure of galaxies, but still coarse enough to necessitate the use of subgrid models for small-scale stellar and black hole processes.

It is currently computationally infeasible to run simulations at this resolution in substantially larger volumes\footnote{The \textsc{BlueTides} simulation \citep{feng_bluetides_2016} is one of the largest high-resolution hydrodynamic simulations, with a volume 400 $h^{-1} \, \mathrm{Mpc}$ on a side, but was only run to $z = 7$.}.
This is an issue for certain science questions, since the volumes typically simulated ($\sim (100 \; \mathrm{Mpc})^3$) do not contain large numbers of rare, overdense environments, as well as galaxies with unusual growth histories (e.g. star bursts).
For example, \eagle\ contains only seven clusters at $z = 0$, and these are all relatively low mass ($ M_{200,\mathrm{crit}} \,/\, \mathrm{M_{\odot}} < 10^{14.5}$).
In order to simulate rare environments, that are not represented in smaller scale periodic volumes, another approach is to use `zoom' simulations \citep{katz_hierarchical_1993,tormen_structure_1997}.
These use initial conditions selected from a much larger dark matter only (DMO) simulation, of order $\sim (1 \; \mathrm{Gpc})^3$ in volume, and then resimulate a smaller region from this volume with full hydrodynamics.
Large scale tidal forces are preserved by simulating the rest of the volume with low resolution dark matter only particles.
This approach has been used successfully to simulate cluster environments with the \eagle\ code \citep{barnes_cluster-eagle_2017,bahe_hydrangea_2017}.

However, since zooms only simulate a small region of interest they have a number of drawbacks compared to periodic simulations.
They cannot be used to predict mean distribution functions \textit{directly}, since they are, by construction, biased regions.
One means of circumventing this limitation is to use multiple zoom simulations of differing environments, and weight the relative abundance of each simulation based on its relative total matter overdensity.
This technique was first demonstrated with the \textsc{GIMIC} simulations \citep{crain_galaxies-intergalactic_2009}, and recently used in the \flares\ simulations to make predictions for the abundance of galaxies during the epoch of reionisation \citep{lovell_first_2021,vijayan_first_2021}.
Another drawback is that zooms cannot be used to self-consistently predict aspects of the large scale structure, such as the clustering of galaxies, since they are by construction non-representative, small volume regions of the Universe.
Large periodic volumes are the only means of studying these kinds of spatial statistics \citep[\textit{e.g.} the BAHAMAS project;][]{mccarthy_bahamas_2017}, but these large volumes cannot currently be simulated at the high resolution necessary to model internal galaxy structures.
This limits what can be achieved with high-resolution hydrodynamic simulations.

$N$-body DMO simulations predict the distribution of matter as a result of gravitational interactions only, and are therefore significantly cheaper computationally than simulations including the gas hydrodynamics.
They are therefore less demanding to run accurately in large volumes, allowing them to be used to explore the large scale structure (LSS).
There are also a number of approaches to modelling galaxy evolution that are relatively simpler than running a full hydro simulation, using semi-analytic or phenomenological models to populate haloes in DMO simulations.
The host halo has a significant impact on the properties of a galaxy; haloes are the cradles within which galaxies form, and continue to influence the evolution of a galaxy throughout its lifetime \citep{wechsler_connection_2018}.
Understanding the relationship between galaxy properties and the properties of their host haloes is an important factor in understanding galaxy formation and evolution, and in the subsequent building of these kinds of galaxy evolution models.

Semi Analytic Models (SAMs) explicitly assume a close relationship between a galaxy and its host-halo.
They treat the complicated physics of galaxy formation with approximate, physically-motivated analytical models, applied \textit{ex post facto} to N-body dark matter only simulations \citep[for a review, see][]{baugh_primer_2006}.
The halo properties, and their merging history, provide the input parameters for such models, which have successfully reproduced a number of distribution functions simultaneously \citep[\textit{e.g.}][]{gonzalez-perez_new_2014, henriques_galaxy_2015, henriques_l-galaxies_2020}.
Subhalo Abundance Matching (SHAM) models also rely explicitly on the galaxy-halo relationship, populating dark matter haloes from simulations with rank ordered galaxies from observations.
Such models have been used to constrain the stellar mass - halo mass relation \citep[\textit{e.g.}][]{behroozi_comprehensive_2010,moster_constraints_2010,moster_galactic_2013,legrand_cosmos-ultravista_2019}, though it has been noted that the efficacy of such methods is highly dependent on the observational selection function \citep{stiskalek_dependence_2021}.
Both these approaches are capable of modelling galaxy evolution over very large volumes, allowing predictions for the clustering of galaxies as well as their evolution in rare, overdense environments.
They have also been used in combination with hydrodynamic simulations in order to highlight potential issues \citep[\textit{e.g.} for satellites where mergers lead to mass loss;][]{simha_testing_2012}, and SAMs have even been explicitly calibrated to reproduce hydrodynamic simulations \citep{neistein_hydrodynamical_2012,mitchell_how_2021}, allowing an investigation into the effects of changes to specific coefficients in the model.

Machine learning methods continue to grow in popularity in all areas of astronomy \citep[see][]{ball_data_2010,fluke_surveying_2020}, and a number of recent papers have explored how they can be used in combination with simulations to emulate galaxy properties, analogous to a SHAM or SAM model.
In a pioneering paper, \cite{xu_first_2013} used the Millennium simulation, coupled with a SAM, to predict the number of galaxies in a given halo using support vector machines and \textit{k}-nearest neighbour algorithms.
Later, \cite{kamdar_machine_2016} showed how tree based methods can be trained to learn additional properties of the the baryon-halo relationship directly from an existing SAM.
They used dark matter properties from each halo as features, and baryonic properties as predictors, and trained the machine to learn the mapping between the two.
They then followed this up by applying the same technique to the \textsc{Illustris} hydrodynamic simulation \citep{kamdar_machine_2016-1}.
\cite{agarwal_painting_2018} presented a similar model applied to the \textsc{MUFASA} simulation.
Using the more recent \textsc{Illustris-TNG} simulation, \cite{jo_machine-assisted_2019} presented a similar model, and then applied this trained model to the much larger DMO MultiDark-Planck simulation.
A novel addition to their model was historical halo features (extracted from the halo merger tree), which allowed the model to broadly reproduce key distribution functions, though we note that they do not present tests in the high halo mass regime ($> 10^{14} \; \mathrm{M_{\odot}}$).
\cite{sullivan_using_2018} used artificial neural networks to better predict the baryon fraction of haloes at high-redshift using both dark matter and baryonic properties from their \textsc{Ramses-RT} radiative transfer simulations.
Most recently, a number of hybrid approaches have been presented: \cite{moews_hybrid_2020} combined the results of an equilibrium model with machine learning on the \textsc{Simba} simulations, and \cite{hearin_generating_2020} combined empirical modelling with simulation outputs from a SAM to populate large DMO volumes with galaxies.
\cite{icaza-lizaola_sparse_2021} demonstrated, using a sparse regression approach, that halo angular momentum has little impact on the stellar-halo mass relation.
Finally, a number of approaches have demonstrated predictions for baryonic properties of the cosmic web not necessarily linked to discrete subhaloes \citep[\textit{e.g.}][]{sinigaglia_bias_2020}.

In this paper we build on these previous works, by combining the results of both periodic and zoom cosmological simulations from the \eagle\ project to train a machine learning model to learn the relationship between galaxy baryonic properties and their host dark matter haloes.
Our approach is unique in two ways.
Firstly, we match subhaloes from each hydrodynamic simulation with those in a DMO counterpart (simulated from the same initial conditions), in order to avoid the effect of baryons on the host dark matter halo \citep{schaller_baryon_2015}.
This allows the model to be directly applied to an independent DMO simulation, without leading to biases in the predictions due to differences in the dark matter features.

Secondly, we address the issue of \textit{generalization error}.
Machine learning methods are a powerful set of techniques for making predictions on data that look similar to the data on which they are trained, but fail when presented with new data that lies outside of the bounds of the original training data.
This presents a problem for models trained on smaller periodic volumes, since such volumes will not contain the massive clusters present in larger DMO simulations, and hence any model trained on these volumes won't provide good predictions for galaxies in overdense environments.
We avoid this by including clusters from the \ceagle\ project \citep[\ceagle;][]{barnes_cluster-eagle_2017,bahe_hydrangea_2017} in our training set.
This allows us to apply the trained model to the much larger volume $(800 \; \mathrm{Mpc})^3$ \pmill\ simulation \citep{baugh_galaxy_2019}, and predict distribution functions of key baryonic properties within this enormous volume, extending the dynamic range, as well as allowing predictions of clustering statistics on larger scales / for higher mass haloes.
The method is shown diagrammatically in \fig{ML_paper_figure}.

Whilst often negatively perceived as a `black box', many machine learning methods in fact provide a wealth of insights into the form of their predictive model, and the weight given to their input parameters.
This presents an opportunity to learn, in an unbiased manner, what parameters best explain the galaxy-halo connection.
We train the model with a range of dark matter properties, and explore the relative predictive power of each one on the baryonic properties.
Hydrodynamic simulations represent the cutting edge of cosmological modelling; machine learning methods could provide a practical way of extracting quantitative information on the modelled relationships.
All of these insights can be used to inform future analytic, semi-analytic and hydrodynamic model development.

This paper is laid out as follows.
In \sec{sims} we present the simulations used to train the model, as well as our algorithm for matching subhaloes between the hydro and DMO runs.
\sec{ml_methods} details the machine learning methods used, as well as our choice of features and predictors.
\sec{test_results} details our results on test sets, including the effect of including density information.
\sec{pmill_predictions} presents our results on independent DMO simulations, including the P-Millennium simulation, and \sec{feature_exploration} shows our feature exploration analysis.
Finally, in \sec{conc} we discuss our results and summarise our conclusions.
Throughout, we assume a (flat) Planck year 1 cosmology \citep[$\Omega_{\mathrm{m}} = 0.307$, $\Omega_{\Lambda} = 0.693$, $h = 0.6777$, ][]{planck_collaboration_planck_2014} and a Chabrier stellar initial mass function \citep[IMF;][]{chabrier_galactic_2003}.

\section{Simulations}
\label{sec:sims}

\subsection{The \eagle\ \& \ceagle\ simulations}
\label{sec:eagle_sim}

\begin{figure*}
	\includegraphics[width=\textwidth]{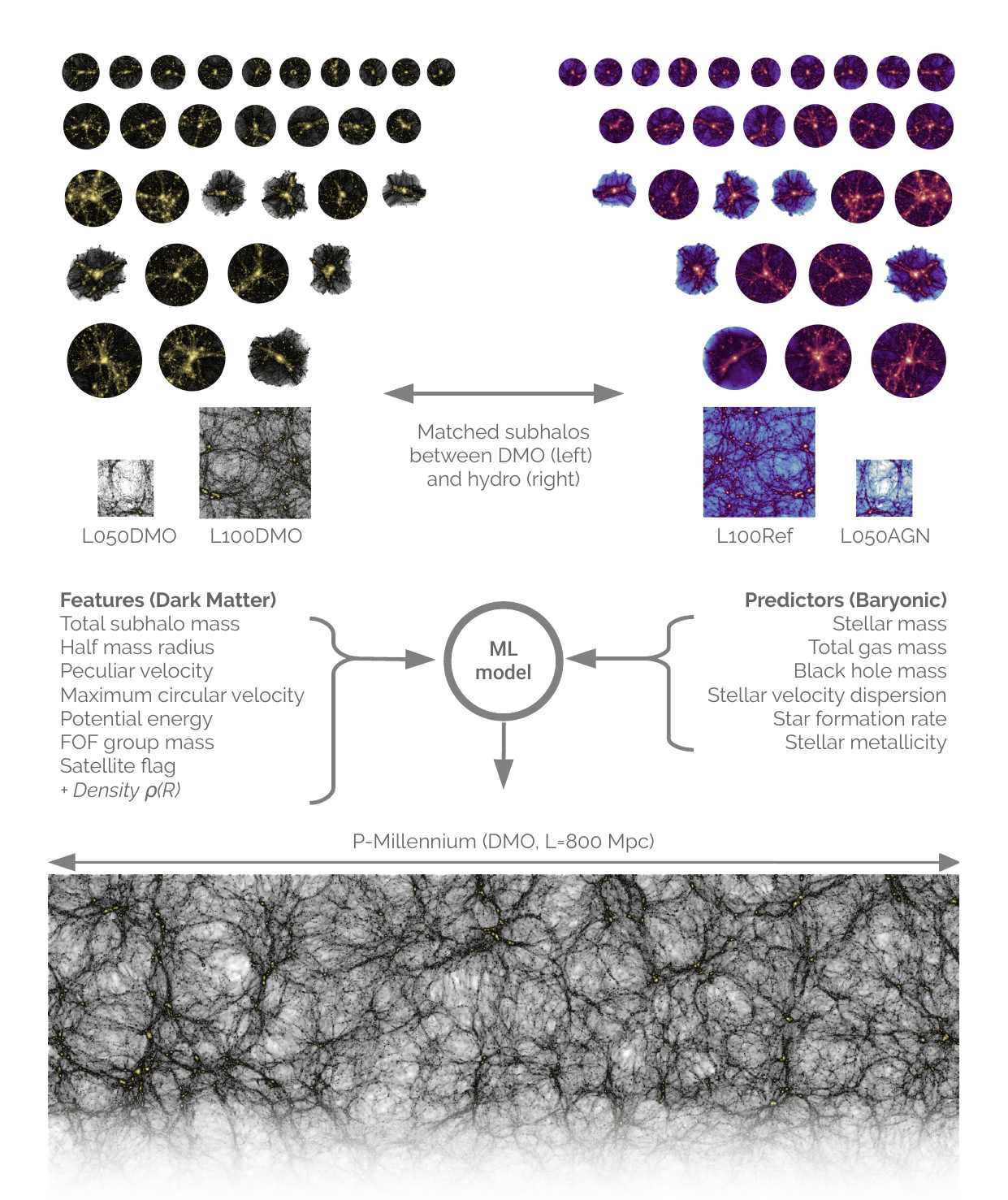}
    \caption{Diagram showing the simulations (approximately to scale) used throughout this work, and the features and predictors used for training the machine learning model.
		At the top are the \ceagle\ zoom simulations; each image shows the distribution of dark matter (left) in the DMO simulations, and the gas (right) in the full hydro simulation, centred on the centre of potential of the most massive FOF group in each simulation, within a radius $r = 15 \times R_{\mathrm{crit},200}$.
		Below these are the cubic periodic \texttt{L100Ref} and \texttt{L050AGN} simulations, again showing the dark matter (left) and gas (right).
		In the centre are tables detailing the features from the DMO simulations (left) and the predictors from the hydro simulations (right).
		At the bottom is a cropped image of the dark matter distribution in the P-Millennium simulation, to which the trained machine learning model is applied to predict the baryonic properties of its haloes.
		}
    \label{fig:ML_paper_figure}
\end{figure*}

\begin{table*}
	\label{tab:simulations}
	\begin{tabular}{ccccccccc}
	Simulation & Prefix & Volume ($\mathrm{Mpc}^3$) & $N_{\mathrm{halo}} (> 10^{10} \, \mathrm{M_{\odot}})$ & $N_{\mathrm{matched}}$ & $N_{\mathrm{train}}$ & $N_{\mathrm{test}}$ & $C_{\mathrm{visc}}$ & $\Delta T$ \\
	\hline
	Reference L0100N1504 & \texttt{L100Ref} & $100^{3}$ & $88\,173$ & $86\,861$ & $69\,615$ & $17\,246$ & $2 \pi$ & $10^{8.5}$ \\
	AGNdT9 L0050N0752 & \texttt{L050AGN} & $50^{3}$ & $11\,423$ & $11\,265$ & $9\,031$ & $2\,231$ & $2 \pi \times 10^2$ & $10^{9}$ \\
	\ceagle\ & \texttt{ZoomAGN} & $202.7^{3}$ & $373\,275$ & $364\,408$ & - & - & $2 \pi \times 10^2$ & $10^{9}$ \\
	\ceagle\ + L050AGN & \texttt{L050AGN+ZoomAGN} & $203.7^{3}$ & $384\,698$ & $375\,673$ & $300\,770$ & $74\,903$ & $2 \pi \times 10^2$ & $10^{9}$ \\
	\end{tabular}
	\caption{Details on each simulation set. The columns provide (1) the name or description of the simulation set, (2) the prefix used throughout this paper, (3) the total volume, (4) the number of subhaloes with mass $> 10^{10} \, \mathrm{M_{\odot}}$, (5) the number of those haloes matched between the hydro and DMO simulations (see \sec{matching}), (6) the number of subhaloes in the training set, (7) the number of subhaloes in the test set, (8) the value of the viscosity parameter, and (9) the value of the $\Delta T$ parameter.}
\end{table*}

\begin{figure*}
	\includegraphics[width=\textwidth]{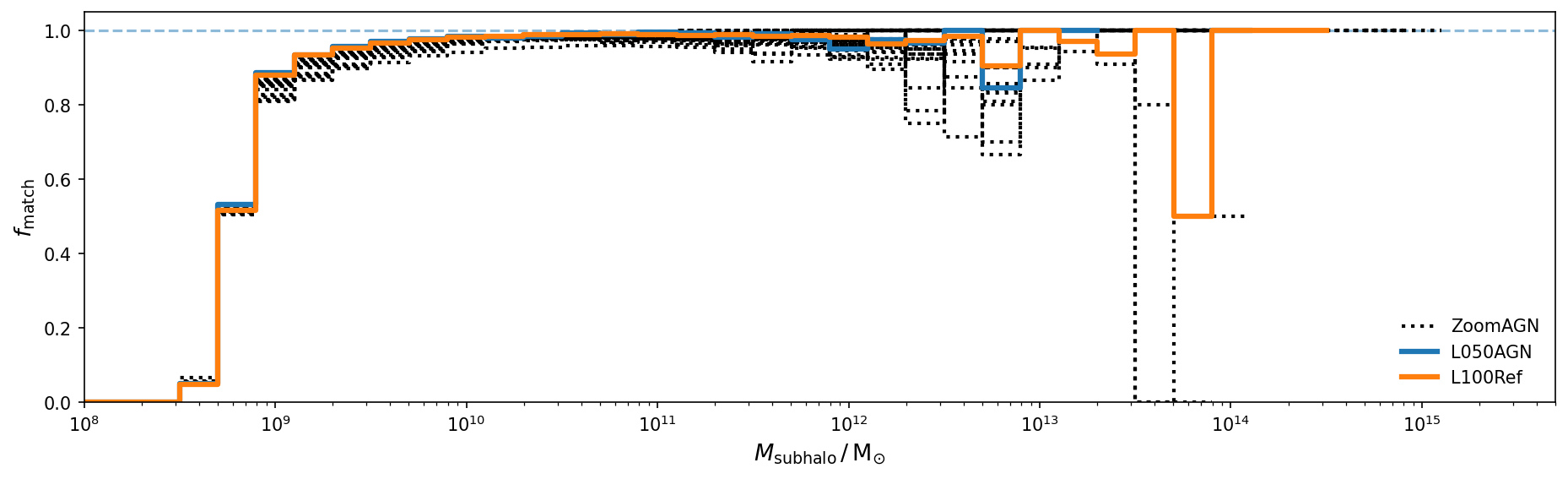}
    \caption{Fraction of subhaloes from each DMO simulation matched with a counterpart in the hydro simulation, binned by total subhalo mass. \texttt{L050AGN} and \texttt{L100Ref} are shown in blue and orange, respectively, and each zoom from \texttt{ZoomAGN} is shown as a black dashed line.
		}
    \label{fig:match_statistics}
\end{figure*}

The \eagle\ project is a suite of cosmological hydrodynamic simulations \citep{schaye_eagle_2015,crain_eagle_2015} employing subgrid models for feedback from stars and AGN.
\eagle\ has been shown to accurately reproduce many observed relations, including the galaxy stellar mass function, galaxy sizes, quenched fractions, gas content and black hole masses \citep{trayford_colours_2015,trayford_optical_2017,lagos_molecular_2015,bahe_distribution_2016,crain_eagle_2017,furlong_size_2017,mcalpine_link_2017} at a range of redshifts \citep[\textit{e.g.}][]{furlong_evolution_2015}.
A number of different resolutions and volumes make up the \eagle\ simulation suite.
In this work we use the `fiducial' resolution simulations, with gas particle mass $m_{g} = 1.8 \times 10^6 \; \mathrm{M_{\odot}}$, dark matter particle mass $9.7 \times 10^{6} \; M_{\odot}$, and a physical softening length of $0.7 \; \mathrm{kpc}$.
haloes in the simulation are identified first through a Friends-Of-Friends (FOF) halo finder, and then split into child self-bound objects with SUBFIND \citep{dolag_substructures_2009}.
Cluster-Eagle \citep[or \ceagle,][]{barnes_cluster-eagle_2017,bahe_hydrangea_2017}, uses the \eagle\ model to simulate cluster environments using the `zoom' re-simulation technique \citep{katz_hierarchical_1993, tormen_structure_1997}.
30 clusters at $z=0$ (shown in \fig{ML_paper_figure}), with a range of halo masses ($14 < \mathrm{log_{10}(M_{200} \,/\, M_{\odot})} < 15.51$), are selected from a (3.2 Gpc)$^3$ `parent' DMO simulation \citep{barnes_redshift_2017}.
The clusters are resimulated at an identical resolution to the fiducial periodic \eagle\ simulation.
Full details on the selected clusters is provided in \cite{barnes_cluster-eagle_2017}.

\ceagle\ uses the AGNdT9 calibration of the \eagle\ model \citep{schaye_eagle_2015}, which, compared to the fiducial Reference model, uses a higher value for $C_{\mathrm{visc}}$, which controls the sensitivity of the BH accretion rate to the angular momentum of the gas, and a higher gas temperature increase from AGN feedback, $\Delta T$.
A larger $\Delta T$ leads to fewer, more energetic feedback events, whereas a lower $\Delta T$ leads to more continual heating.
\cite{schaye_eagle_2015} show that AGNdT9 predicts X-ray luminosities and hot gas fractions in galaxy groups in better agreement with observational constraints, though with some discrepancies on cluster scales \citep{barnes_cluster-eagle_2017}.

\tab{simulations} details the simulations used in this work, and any combinations.
\texttt{L100Ref} is a $(100 \, \mathrm{Mpc})^3$ periodic volume (shown in \fig{ML_paper_figure}) run with the Reference model parameters; the hydro simulation contains $1504^3$ dark matter and $1504^3$ gas particles.
\texttt{L050AGN} is a smaller, $(50 \, \mathrm{Mpc})^3$ periodic volume (shown in \fig{ML_paper_figure}) run at the same resolution as \texttt{L100Ref} but with the AGNdT9 model parameters; it contains $752^3$ dark matter and $752^3$ gas particles.
\texttt{L050AGN+ZoomAGN} is a combination of \texttt{L050AGN} with the zoom cluster regions from \ceagle.
We also match with DMO counterparts to each of these simulations, run using the same initial conditions; the match is described in \sec{matching}.
We use the snapshot corresponding to $z = 0.101$ in all simulations.

Throughout the rest of this text, whenever we refer to a \textit{model} we are referring to a \textit{machine learning} model (unless otherwise stated) trained on the matched hydro-DMO simulations indicated in the name.
The simulations are all referred to explicitly as \textit{simulations} to distinguish them from the machine learning models.

\subsection{Matching between hydrodynamic and dark matter only simulations}
\label{sec:matching}

Including baryons can lead to significant alterations to the underlying dark matter haloes \citep{weinberg_baryon_2008}.
For example, \cite{schaller_baryon_2015} demonstrate that, in the \eagle\ simulation, the halo centres are more `cuspy' in the presence of stars.
In order to apply our trained model to DMO simulations it is necessary to avoid these effects, as they will bias any predictions based on the dark matter features.
We achieve this by matching subhaloes in each hydrodynamic simulation to their counterparts in DMO simulations, and use the properties of the matched haloes in the DMO simulation as our features.
The galaxy properties that a given halo would have if hydrodynamics had been included are then predicted.
Each DMO simulation is run from the same initial conditions, but is not split into baryonic and dark-matter species.
Aside from this, all cosmological and numerical parameters are identical.

We perform the match using the approach of \cite{schaller_baryon_2015}.
We first find the 50 most bound dark matter particles in a subhalo in the hydro simulation, and search for haloes in the \textsc{DMO} simulation that have 50\% or more of these same particles (matched on particle ID).
We then perform the same match in reverse (subhaloes in the DMO matched with subhaloes in the hydro simulation).
Those haloes that match bijectively are linked.

\fig{match_statistics} shows the fraction of haloes matched from the DMO simulation at a given DMO halo mass for the two periodic simulations (\texttt{L100Ref} and \texttt{L050AGN}) as well as each of the \ceagle\ clusters.
We also detail the total number of haloes and the number of matched haloes for each simulation set in \tab{simulations}.
More than 95\% of subhaloes with $M_{\mathrm{subhalo}} > 10^{10} \, \mathrm{M}_{\odot}$ are matched bijectively across all simulations.
We hence choose to train our model only on subhaloes with masses above this threshold (see \sec{training} for details).
By using a threshold dependent only on the DMO properties we can use a similar threshold in any target DMO simulation (subject to the existing resolution constraints of that simulation).

It is noticeable that there are a larger fraction of subhaloes at the high-mass end ($M_{\mathrm{subhalo}} \,/\, M_{\odot} > 5 \times 10^{12}$) that are not matched, in both the periodic and zoom simulations.
We looked at these cases individually, and found, where a single halo was identified in the DMO simulation, the halo finder splits this halo into multiple individual haloes in the baryonic simulation.
Missing these haloes reduces the size of our training set, which is particularly disappointing at the high-mass end where the number of haloes is already low, however we do not expect this to lead to biases in our predictions due to the already heterogenous nature of our training set.

\subsection{The \pmill\ simulation}
\pmill\ is a large DMO simulation ($800 \; \mathrm{cMpc}$ on a side; particle mass $1.06 \times 10^8 \, \mathrm{M_{\odot}}$) using the same \cite{planck_collaboration_planck_2014} cosmology as \eagle.
\cite{baugh_galaxy_2019} first presented the simulation, and demonstrated its use as a parent volume for the \galform\ model, in order to predict the atomic hydrogen content of galaxies.
\cite{safonova_rosella_2020} also used \pmill\ as a parent simulation for a SHAM model, generating mock catalogues.
\pmill\ uses the same FOF and Subfind structure finders as the \eagle\ simulation project, which means the features can be used directly for any model trained on \eagle.
We present our predictions using \pmill\ in \sec{pmill_predictions}.

\section{Machine Learning Methods}
\label{sec:ml_methods}

\subsection{Extremely Randomised Trees}
\label{sec:ert}

We used the \textsc{Scikit-Learn} \citep{pedregosa_scikit-learn:_2011} implementation of Extremely Randomised Trees \citep[ERT;][]{geurts_extremely_2006}, a tree based ensemble method.
ERT is demonstrably effective in this domain compared to other popular machine learning methods \citep{kamdar_machine_2016,jo_machine-assisted_2019}.

To understand what makes ERT such an effective learner, first consider a single decision tree.
Decision trees are typically constructed top down, numerically evaluating all splits for each feature using a cost function.
The best split (lowest cost) is chosen at each level.
Some of the issues seen with Decision Trees, particularly overfitting, can be alleviated by ensembling many different trees trained on subsets of the data.
Random Forests extend this idea by, at each split, randomly limiting the feature space from which splits can be made (within individual trees not all of the data is used, but over the whole ensemble it is).
This increases the variance by stopping strong features from dominating each tree.
ERT also introduces another layer of randomness; each split is chosen at random from the range of values available for each feature.
Bad splits are still rejected, but the extra layer of randomness encourages exploration of the full feature space, creating more `weak' learners for use in the ensemble.
At each iteration, only the best split from the subset of features is chosen, and the iterative procedure continues until a leaf node condition is reached.

Within ERT the mean squared error (MSE) is used to evaluate each split.
To quantify the effective fit of each model, and to discriminate between models, we used both MSE and the Pearson correlation coefficient ($\rho$), defined below:

\begin{equation}
  \rho = \frac{\mathrm{cov} (X_{\mathrm{predicted}} X_{\mathrm{test}})}{\sigma_{X_{\mathrm{predicted}}} \sigma_{X_{\mathrm{test}}}}  \;\;.
\end{equation}

\subsection{Features \& Predictors}
\label{sec:features_predictors}

\begin{figure*}
	\includegraphics[width=\textwidth]{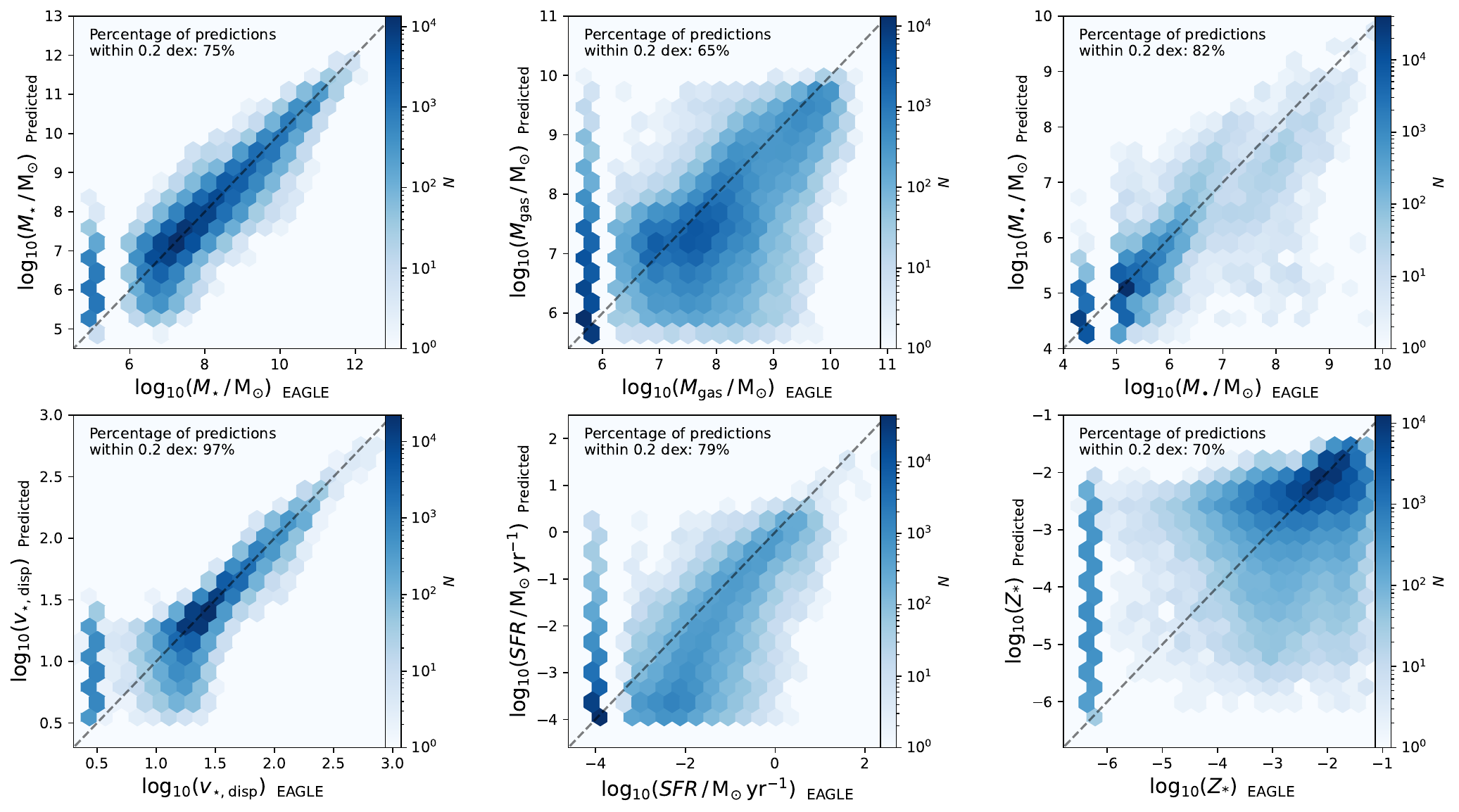}
    \caption{Predicted (from the machine learning model) against the true baryonic properties on the test set from the \texttt{L050AGN+Zoom} simulation set.
    Clockwise from top left: stellar mass, gas mass, black hole mass, star formation rate, stellar metallicity, and stellar velocity dispersion.
    The vertical bar separated from the rest of the distribution to the left in each panel corresponds to galaxies with a true value of zero for that corresponding predictor (see \sec{features_predictors}).
    The fraction of galaxies whose predicted property is within 0.2 dex of the true value is quoted at the top left of each panel.
    }
    \label{fig:joint_plots_L0050N0752}
\end{figure*}


We chose our features from the properties of the DMO haloes and their host FOF haloes.
Some features are expected to be of greater importance for predicting certain baryonic properties; we explore this in \sec{feature_exploration}.
The selected subhalo features are as follows: total subhalo mass ($M_{\mathrm{sub}}$), half mass radius ($R_{1/2}$), peculiar velocity ($v$), maximum circular velocity ($v_{\mathrm{max}}$), radius of maximum circular velocity ($R_{v_{\mathrm{max}}}$), potential energy ($E_{\mathrm{p}}$), FOF group mass ($M_{\mathrm{crit,200}}$), and finally a boolean feature that specifies whether the subhalo is a satellite or a central.

Since we wish to evaluate the impact of environment we also include additional features to quantify this.
As a simple measure of environment we calculated the density of dark matter within spheres centred on a given subhalo in the DMO simulation.
We ran a periodic KD-tree search for neighbouring particles, then calculated the density on different scales, $R = [1,2,4,8] \, \mathrm{Mpc}$, to quantify both the small and large scale environment.
We indicate in the text where these additional features are included in a given training set.

More dark matter features are available in the subfind catalogues, and additional features could be calculated from the particle information (such as the large scale tidal torque), but we limited our chosen features to those above as they are present in both the \eagle\ and \pmill\ catalogues.
Combinations of features may also lead to better predictive accuracy; we will explore this systematically in future work.

We predict six baryonic properties: the stellar mass, gas mass, black hole mass, stellar velocity dispersion, star formation rate and stellar metallicity.
The stellar mass and gas mass are taken from the central 30 kpc of each subhalo to allow better comparison with observations.
We transform all of these predictors into log space, which has been shown to improve the prediction accuracy due to the typically large dynamic range of cosmological properties \citep{jo_machine-assisted_2019}.
If the value is zero, we set it to some small value, determined by the resolution limit where appropriate,
\begin{align*}
	M_{\star} \,/\, \mathrm{M_{\odot}} &\geqslant 1 \times 10^5 \\
	M_{\mathrm{gas}} \,/\, \mathrm{M_{\odot}} &\geqslant 5 \times 10^5 \\
	M_{\bullet} \,/\, \mathrm{M_{\odot}} &\geqslant 2 \times 10^4 \\
	\mathrm{SFR} \,/\, \mathrm{M_{\odot} \; yr^{-1}} &\geqslant 1 \times 10^{-4} \\
	\mathrm{Z_{*}} &\geqslant 5 \times 10^{-7} \\
	v_{\star,\mathrm{disp}} \,/\, \mathrm{km \; s^{-1}} &\geqslant 3  \;\;.
\end{align*}

\subsection{Training}
\label{sec:training}

We train our model on all haloes with a dark matter mass (as measured in the \textsc{DMO} simulation) $M_{\mathrm{sub}}\,/\, \mathrm{M_{\odot}} \geqslant 1 \times 10^{10}$.
The completeness of our selection with respect to stellar mass is shown in detail in \app{completeness}.
By applying our selection to the dark matter properties we can use the same thresholds when applying the model to independent dark matter only simulations.
We split our data into training and test sets, 80-20\% respectively.
All hyperparameter optimisation, parameter scaling, and training is done on the training set, and only final model assessment is performed on the test set.
For each feature set, the hyperparameters of the ERT instance are chosen through an exhaustive grid search.
For each set of hyperparameters, \textit{k}-fold cross validation is performed \citep{stone_cross-validatory_1974} with $k = 10$ folds, and the coefficient of determination, $\rm{R}^{2}$, is used to discriminate,
\begin{equation}
	R^{2} = 1 - \frac{\sum_{i} (X^{i}_{\mathrm{test}} - X^{i}_{\mathrm{predicted}})^{2}}{\sum_{i} (X^{i}_{\mathrm{test}} - X_{\mathrm{mean,train}})^{2}} \;\;.
\end{equation}
We standardise all of our features and predictors by subtracting the mean and scaling to unit variance.

\section{Predicting Baryonic Properties from Dark Matter Properties}
\label{sec:test_results}

\begin{figure}
	\includegraphics[width=\columnwidth]{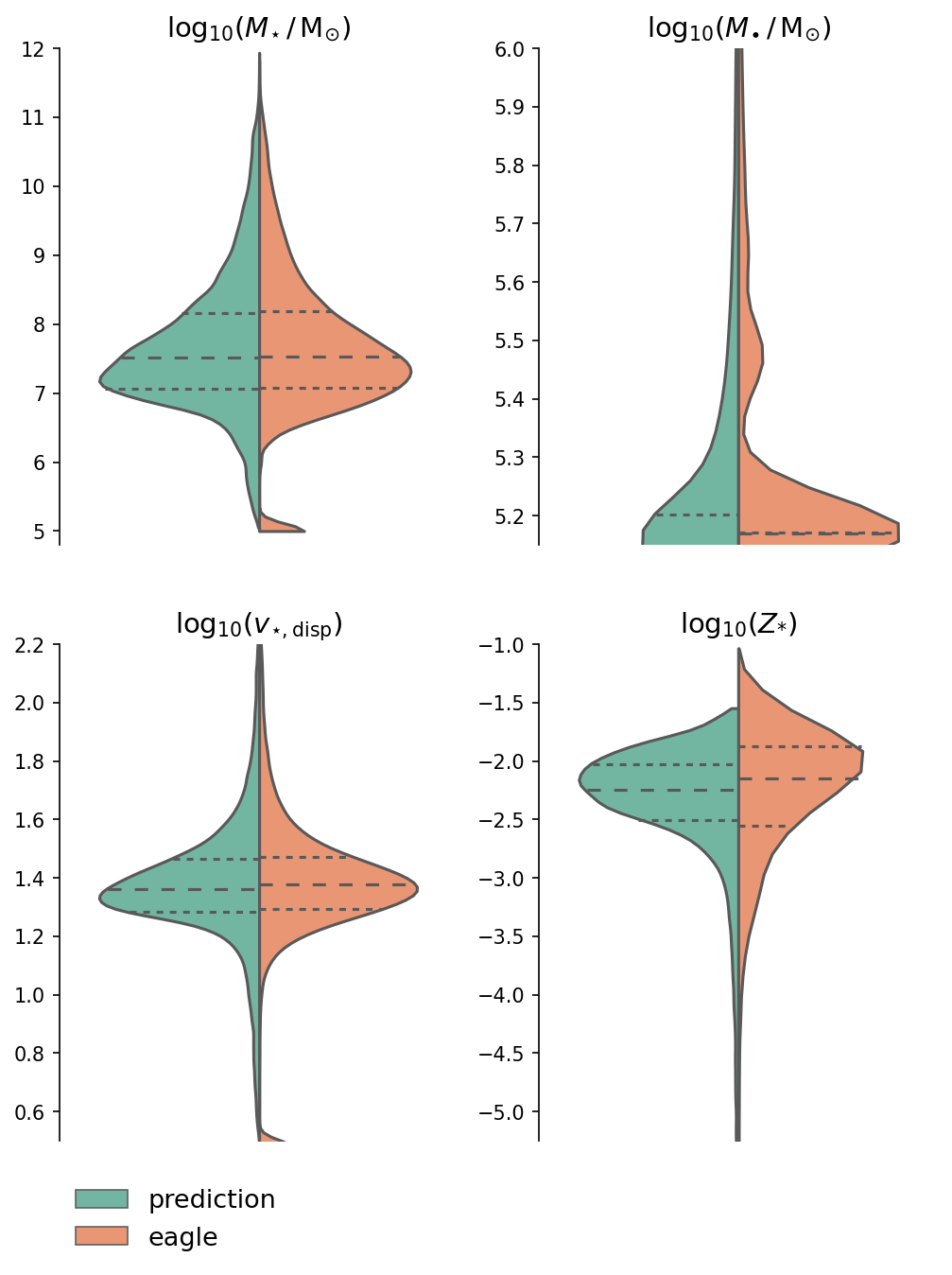}
    \caption{Violin plots showing the distribution of predicted baryonic properties (green) from the machine learning model against the true values (orange) in the \texttt{L050AGN+Zoom} simulation set.
		Dashed and dotted lines show the median and upper/lower quartiles of each distribution, respectively.
		Each distribution is a kernel density estimate of the true underlying distribution, which may smooth some features, particularly where the distribution is discontinuous (\textit{e.g.} galaxies with zero gas mass).
		Clockwise from top left: stellar mass, black hole mass, stellar velocity dispersion, and stellar metallicity.
		}
    \label{fig:violins_L0050N0752}
\end{figure}

\begin{figure}
	\includegraphics[width=\columnwidth]{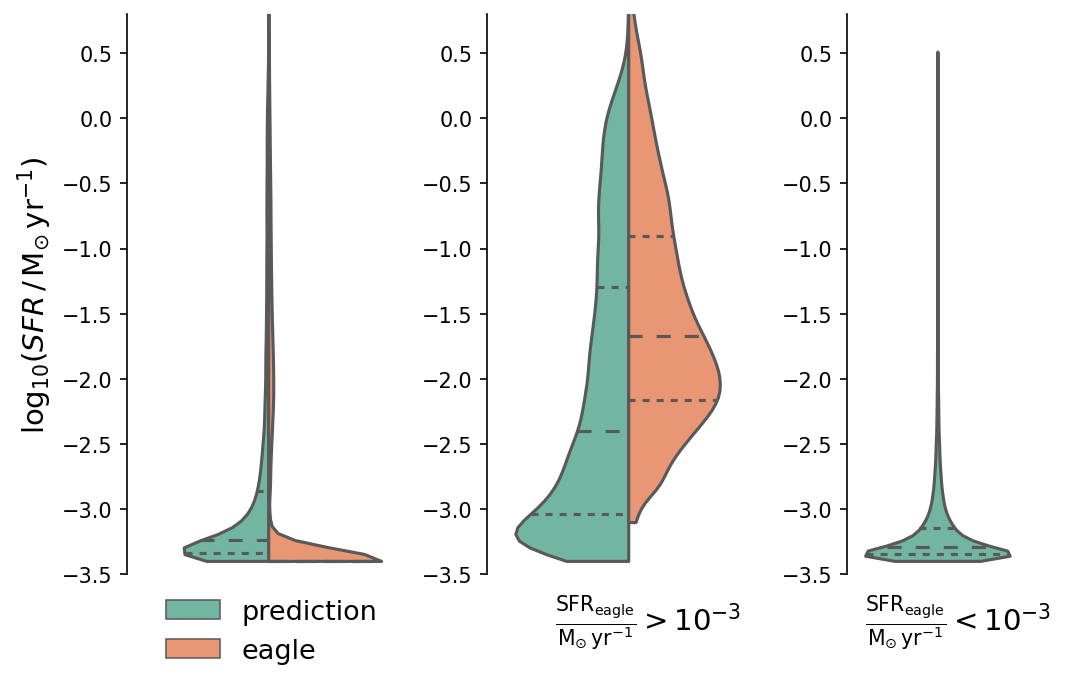}
    \caption{Violin plots showing the distribution of predicted SFR (green) from the machine learning model against the true SFR (orange) in the \texttt{L050AGN+Zoom} simulation set.
		The left plot shows the total distribution, which is heavily skewed toward quiescent galaxies, since the sample is dominated by low mass galaxies that are artificially quenched.
		The central plot shows the distribution ignoring those galaxies with zero SFR in the test set.
		The right plot shows only the \textit{predicted} SFR for all galaxies with zero SFR in the test set (note that this violin is symmetric as only a single property is plotted)).
		}
    \label{fig:violins_SFR_L0050N0752}
\end{figure}

\begin{figure}
	\includegraphics[width=\columnwidth]{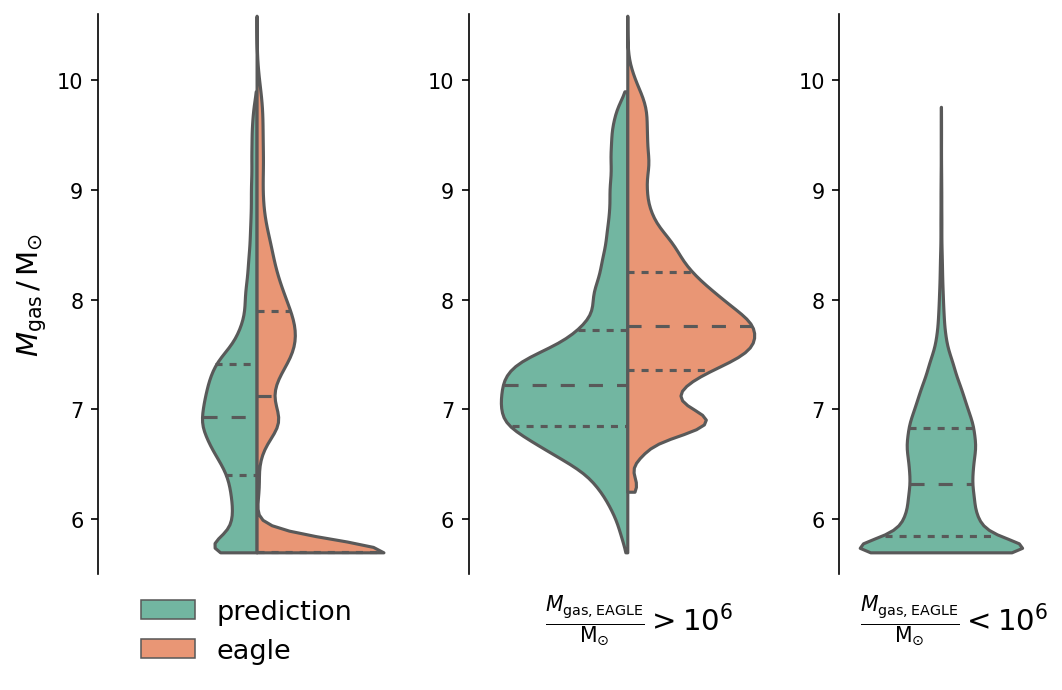}
    \caption{As for \fig{violins_SFR_L0050N0752}, but showing the distribution of total gas mass.
		}
    \label{fig:violins_gas_L0050N0752}
\end{figure}

\begin{figure*}
	\includegraphics[width=\textwidth]{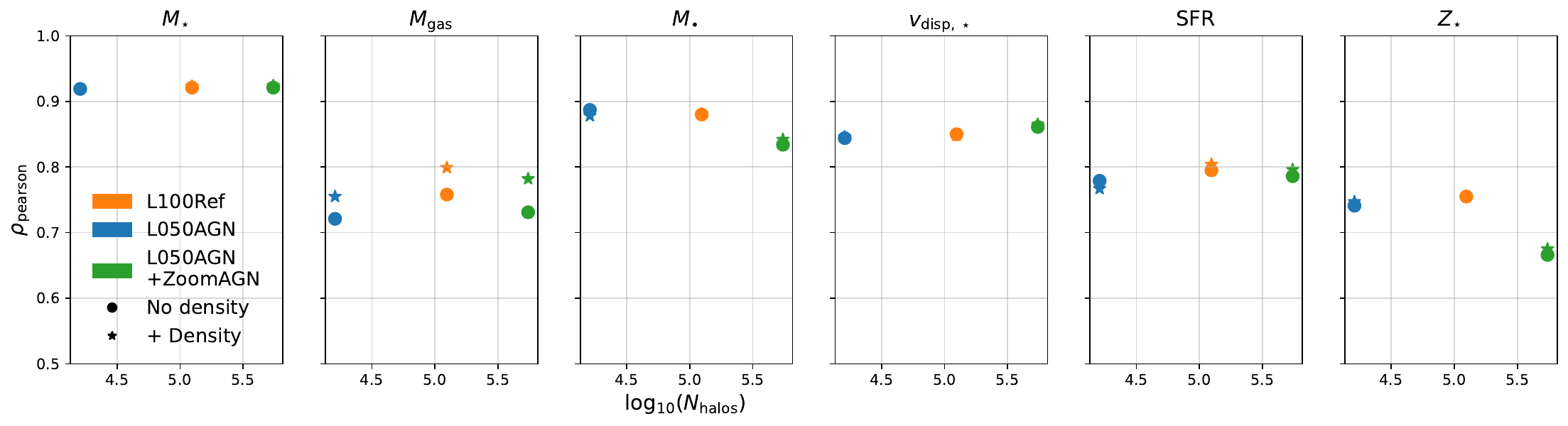}
    \caption{Comparison of fit accuracy described by the Pearson correlation coefficient ($\rho_{\mathrm{pearson}}$), measured on the test set, against the number of haloes in the training set, for each of the baryonic predictors.
		The \texttt{L100Ref}, \texttt{L050AGN} and \texttt{L050AGN+Zoom} simulation sets are shown in orange, blue and green, respectively, where bullet markers show results with the fiducial feature set, and star markers show result including all local density features (see \sec{density}).
		}
    \label{fig:fit_comparison}
\end{figure*}

We first present results for the \texttt{L050AGN+ZoomAGN} model, with the fiducial feature set (excluding environmental features).
\fig{joint_plots_L0050N0752} shows the predicted against the true value for the six baryonic properties in the test set.
Figures \ref{fig:violins_L0050N0752}, \ref{fig:violins_SFR_L0050N0752} \& \ref{fig:violins_gas_L0050N0752}  compare the predicted and true distribution of these properties in the test set as violin plots\footnote{Bin width of the kernel density estimate is calculated using Scott's rule \citep{scott_optimal_1979}.}.
Together, these figures show how accurately the model predictions are, and how well the cosmic distribution is reproduced.
We also quote the fraction of galaxies where the predicted value is within 0.2 dex of the true value; for the stellar velocity dispersion this is as high as 97\%, but even for the gas mass, which has the lowest prediction accuracy, this is still close to two thirds of the sample (65\%).
This is comparable to the accuracy achieved in \cite{neistein_hydrodynamical_2012} in their SAM trained on a hydro simulation, though we push our predictions to lower stellar masses.

To qualitatively demonstrate the accuracy of the ERT model, we compare to predictions utilising a single feature (subhalo mass or $V_{\mathrm{max}}$), analogous to a SHAM approach.
For this single feature we fit an isotonic regression model\footnote{see \href{https://scikit-learn.org/stable/modules/generated/sklearn.isotonic.IsotonicRegression.html}{here} for details on the Isotonic regression model employed.} between the feature and each predictor (for the whole dataset, not just training).
This model ensures monotonicity, and broadly fits each predictor well considering the simplicity of the model.
We again quote the fraction of galaxies where the predicted value from this simple relation is within 0.2 dex of the true value.
The ERT model shows greater accuracy for all predictors compared to the Isotonic model, whether subhalo mass or $V_{\mathrm{max}}$ are used.
This is particularly the case for the gas mass (49\% where $V_{\mathrm{max}}$ is used, compared to 65\% for the ERT model).
Full details on the Isotonic fits, and comparison to the predicted GSMF and projected correlation function, are provided in \app{shamcomp}.

The model is able to accurately predict both the stellar mass and stellar velocity dispersion remarkably well, however there is more structure in the joint plots for other properties.
Predictions for the stellar metallicity show a greater spread than the other values, perhaps unsurprisingly due to its known complex dependence on the star formation history, however the violin plot shows that the overall distribution is recovered.
Black hole masses in \eagle\ are dominated by newly formed black holes at the seed mass ($10^{5} \; M_{\odot}$), as more haloes reach the mass-threshold for black hole seeding.
The model is able to capture these, and does a reasonable job of predicting the masses of more massive black holes.

The relations for the total gas mass and SFR are more complicated.
There are a large number of galaxies with zero star formation, and the right panel of \fig{violins_SFR_L0050N0752} shows that the model predicts a range of SFRs for these galaxies, though the majority are limited to $< 3 \times 10^{-3} \; \mathrm{M_{\odot} \, yr^{-1}}$.
To see how well the model predicts the distribution of star forming galaxies, we show in the middle panel of \fig{violins_SFR_L0050N0752} the distribution of SFR ignoring quiescent galaxies.
It is clear that the model underpredicts the SFR for most galaxies.
This may be due to the quiescent galaxies biasing the predictions for other haloes, as well as ERT predicting a smooth distribution of SFRs when a discontinuous distribution would be more appropriate.
The SFR is also known to be more strongly dependent on the assembly history;  including features that encode this may lead to better predictions, which we discuss in \sec{conc}.

\fig{violins_gas_L0050N0752} shows that, as for the SFR, there is a reasonably tight relation for the total gas mass, except where galaxies have zero gas mass.
These galaxies make up a large proportion of all subhaloes, and the model fails to predict low gas masses for these galaxies, instead predicting a wider range of gas masses, as can be seen in the right panel of \fig{violins_gas_L0050N0752}.
This suggests that the physics that causes the evacuation of gas from low mass haloes is not encoded in the provided dark matter parameters.
However, the overall distribution, when renormalised, better reproduces that seen in the test set compared to the SFR.

To demonstrate the impact of adding the \ceagle\ clusters to our training set, we compare the prediction accuracy against models trained only on the periodic volumes.
\fig{fit_comparison} shows the Pearson correlation coefficient for the \texttt{L100Ref}, \texttt{L050AGN} and \texttt{L050AGN+zoom} models.
Adding the zoom regions leads to a large increase in the training set size, but this has no significant positive effect on the predictive accuracy for any of the features.
In fact, for the gas mass, black hole mass and stellar metallicity the predictive accuracy is actually worse.
This may be due to the unique impact of the cluster environment on these three particular baryonic properties of galaxies, for example through the effect of ram pressure stripping and fly-by interactions.
So while there is more data for the machine to learn from, the relationship represented is more complicated than that present in the periodic volumes, and therefore more difficult to predict.
We stress that in order to make predictions for larger boxes, it is essential to include these environments in the training set, and that a lower predictive accuracy compared to the periodic volumes is not necessarily indicative of a poorer model.

This does not suggest that adding more data does not improve the predictive accuracy - $\rho_{\mathrm{pearson}}$ calculated for \texttt{L100Ref} is higher than than for \texttt{L050AGN} for all baryonic properties, showing the advantage of a larger training set size where the underlying distribution of galaxy properties is broadly similar.

\subsection{The effect of including local density in the feature set}
\label{sec:density}

We add four features for the local density calculated within spheres with radii $R = [1,2,4,8] \, \mathrm{Mpc}$.
\fig{fit_comparison} also shows the impact of including these additional features on the predictive accuracy for the \texttt{L050AGN}, \texttt{L100Ref} and \texttt{L050AGN+Zoom} simulation sets.
Including density information has a minor positive impact on the predictive accuracy of all features for almost all simulation sets, though the quantitative impact is small in most cases.
The largest impact is seen for the gas mass, with an increase in $\rho_{\mathrm{pearson}}$ of approximately $+0.05$ for the periodic simulation sets, and $+0.07$ for the \texttt{L050AGN+Zoom} simulation set.
This fits with the hypothesis suggested above that environmental effects operating in clusters lead to poor predictions for the gas mass when environmental features are not included.
Such features are important for accurately predicting specific baryonic properties.

In summary, our model is capable of predicting a range of baryonic properties with reasonable accuracy, and successfully reproduces their cosmic distributions.
We now show how the model can be applied to independent, larger DMO volumes, and the impact of including the zoom regions on the predicted relations.

\section{Application to DMO simulations}
\label{sec:pmill_predictions}

\begin{figure*}
	\includegraphics[width=\textwidth]{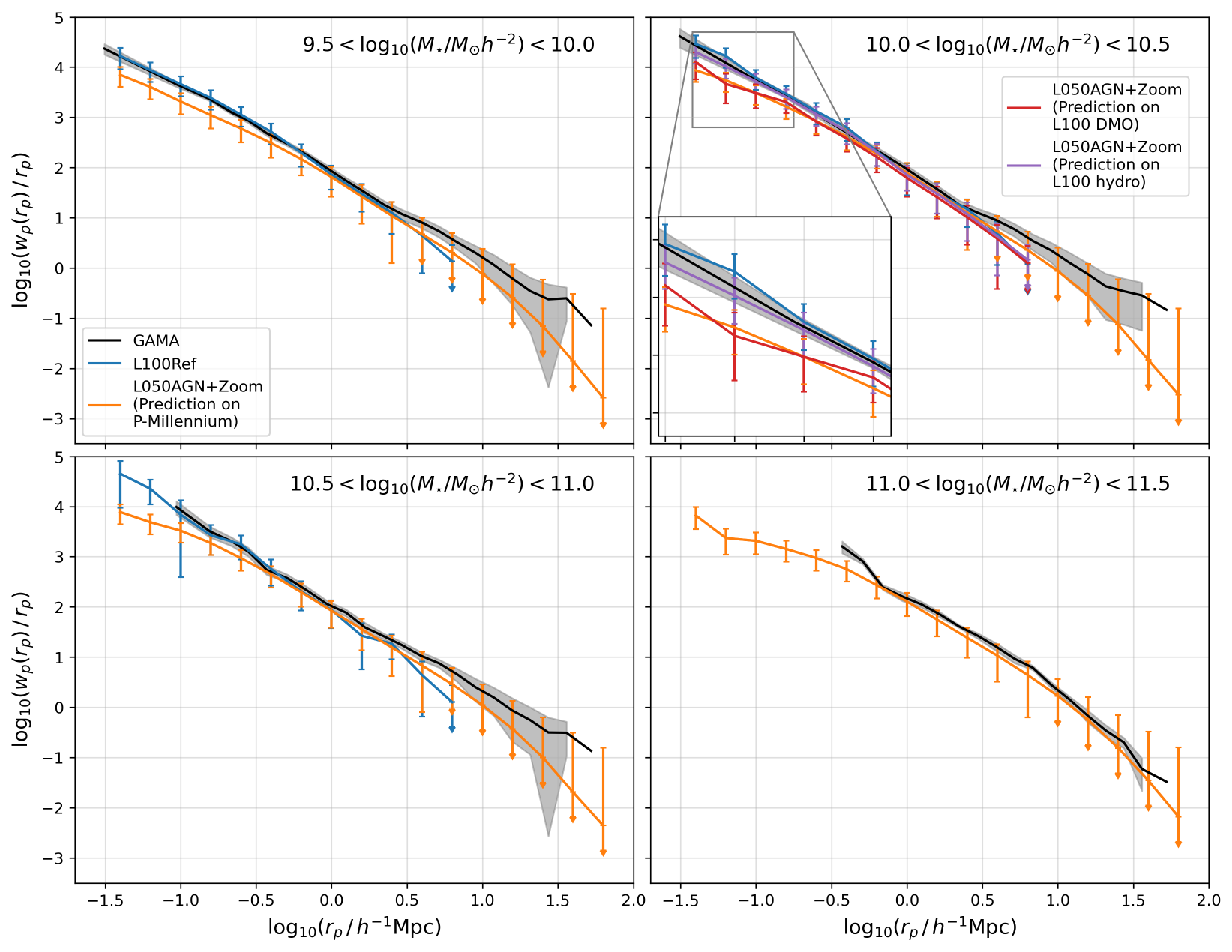}
    \caption{
		Projected correlation function in bins of stellar mass; the mass range is indicated in each column.
		The results from \texttt{L100Ref} are shown in blue, and the \texttt{L050AGN+Zoom} machine learning model predictions on P-Millennium are shown in orange.
		Observational results from GAMA \citep{farrow_galaxy_2015} are shown in grey.
		Errors are estimated using jacknife resampling of each simulation volume.
		}
    \label{fig:clustering}
\end{figure*}

A key aim of the model is to produce predictions for distribution functions and clustering statistics for much larger volumes than can be achieved using periodic hydrodynamic simulations.
To this end we test how well our model produces the two-point galaxy correlation function (2PCF), galaxy stellar mass function (GSMF), star-forming sequence, stellar mass -- metallicity relation and the stellar mass -- black hole relation in independent DMO volumes, including the $(800 \, \mathrm{Mpc})^3$ P-Millennium simulation.

\subsection{The Two-Point Galaxy Correlation Function}
Galaxy clustering measurements provide a powerful means of testing gravity and  cosmological parameters, including the contribution of dark energy, as well as the impact of galaxy bias on galaxy formation and evolution.
One of the key statistics for measuring clustering is the spherically averaged two-point correlation function (2PCF) \citep{peebles_large-scale_1980}, defined as
\begin{equation}
	\xi(r) = \frac{1}{\langle n \rangle} \frac{\mathrm{d}P}{\mathrm{d}V} - 1,
\end{equation}
where $\langle n \rangle$ is the mean comoving number density of galaxies,and  $\mathrm{d}P/\mathrm{d}V$ is the probability of finding a galaxy in volume $\mathrm{d}V$ at comoving distance $r$ from another galaxy.
For redshift surveys, where the line-of-sight distance is inaccessible, this is often split into projected and line-of-sight distance components, which can be used to estimate the \textit{projected} correlation function \citep{davis_survey_1983},
\begin{equation}
	w_{\mathrm{p}} (r_{\mathrm{p}}) = 2 \int^{\pi_{\mathrm{max}}}_{0} \xi(r_{\mathrm{p}}, \pi) \, \mathrm{d} \pi,
\end{equation}
where $\pi_{\mathrm{max}}$ is the maximum distance along the line-of-sight.
Since $w_{\mathrm{p}} (r_{\mathrm{p}})$ is robust against redshift space distortion effects it is better suited for comparisons with simulations.

Simulation studies of galaxy clustering are typically carried out on large scales with DMO simulations or relatively lower resolution hydrodynamical simulations \citep[\textit{e.g.} BAHAMAS;][]{mccarthy_bahamas_2017}, and on smaller scales using high-resolution hydrodynamical simulations, which can resolve the baryonic feedback effects on haloes \citep[see][]{van_daalen_impact_2014}.
We here see how well our machine learning model can provide predictions on both large and small scales \textit{simultaneously} by applying the model to the large-volume P-Millennium simulation.
We estimate errors on our clustering statistics using jacknife resampling of each simulation volume \citep[for details, see][]{artale_small-scale_2017}.

\fig{clustering} shows the projected 2PCF measured on the \texttt{L100Ref} simulation, the \texttt{L050AGN+Zoom} model applied to the P-Millennium simulation, and compared to observational results from GAMA \citep{farrow_galaxy_2015} in different stellar mass bins.
As shown in \cite{artale_small-scale_2017}, the \texttt{L100Ref} simulation is in good agreement with the observational constraints on small scales up to stellar masses of $10^{11} \, \mathrm{M_{\odot}}$.
However, on larger scales ($r_{\mathrm{p}} > 3 h^{-1} \mathrm{Mpc}$) there is a deficit in the normalisation, attributed to finite-volume effects; the smaller periodic boxes do not sample the largest modes in the power spectrum.
There are also too few galaxies above a stellar mass of $10^{11} \, \mathrm{M_{\odot}}$ in \texttt{L100Ref} to obtain robust clustering statistics.

The \texttt{L050AGN+Zoom} model, applied to the much larger volume P-Millennium simulation, shows no such deficit at the largest scales.
We are in fact able to make predictions out to scales of $100 \, h^{-1} \, \mathrm{Mpc}$, a factor of ten larger than achievable with the periodic simulations.
The model is also able to make predictions for the clustering of the most massive galaxies, $> 10^{11} \, \mathrm{M_{\odot}}$, since there are sufficient numbers of these galaxies to produce reliable statistics.

There is, however, a small deficit in the normalisation at the smallest scales in the lower mass bins for the \texttt{L050AGN+Zoom} model (outside the estimated errors).
This may be due to a number of effects, one being the lower resolution of the P-Millennium simulation, which may lead to sub-structures on small scales being smoothed out.
To test the impact of this we applied the \texttt{L050AGN+Zoom} model to the DMO $100 \, \mathrm{Mpc}$ box (using the same initial conditions as the \texttt{L100Ref} simulation), which has a mass resolution $\sim 10 \times$ higher.
This is shown in \fig{clustering}; at the largest scales the model shows the same deficit as the \texttt{L100Ref} simulation, due to the smaller box size.
However, at small scales there is the same deficit as in the \texttt{L050AGN+Zoom} model applied to P-Millennium.
This confirms that it is not resolution effects leading to the lower amplitude.

\begin{figure*}
\begin{multicols}{2}
	\includegraphics[width=\columnwidth]{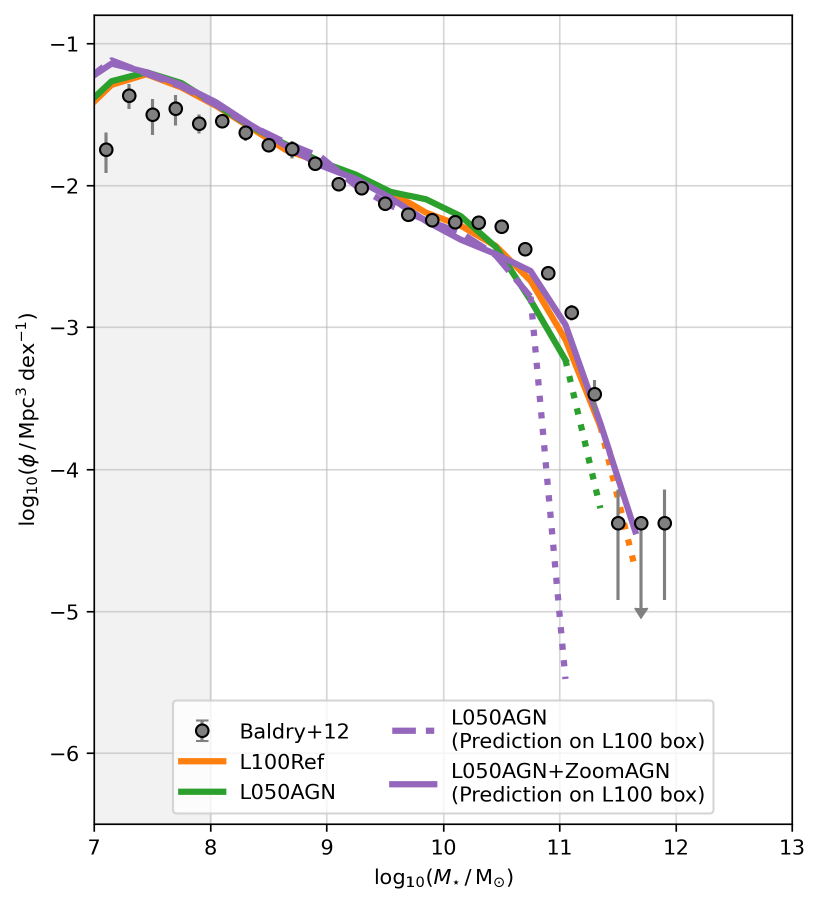} \\
	\includegraphics[width=\columnwidth]{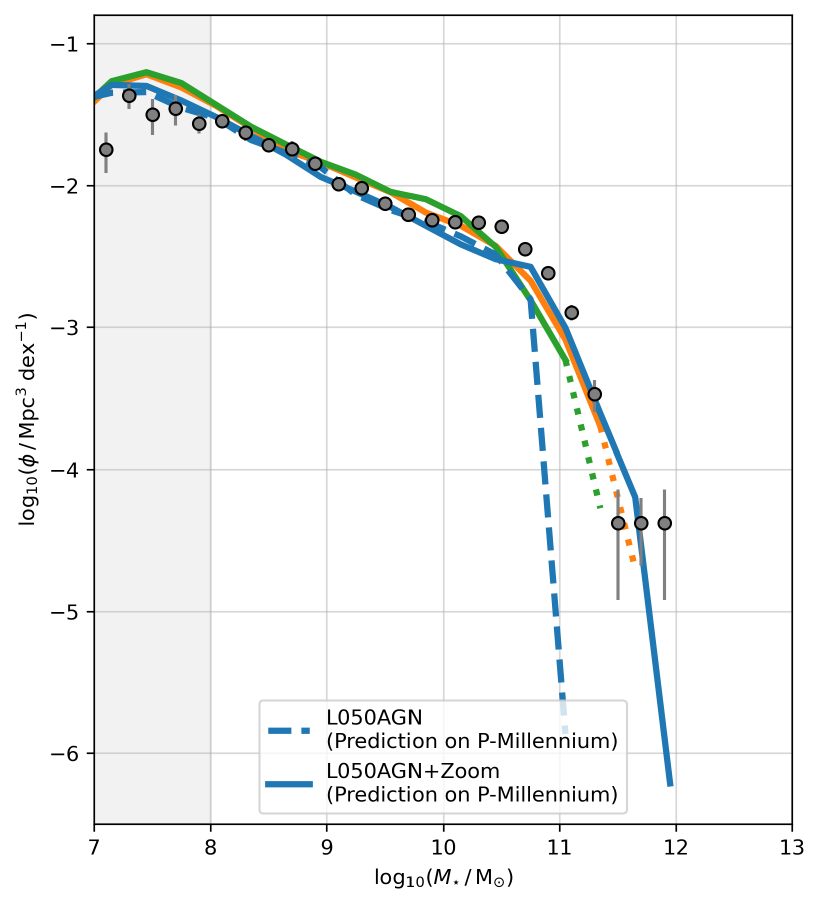}
	\end{multicols}
    \caption{The galaxy stellar mass function (GSMF).
		Both panels show the GSMF from the \texttt{L100Ref} (orange) and \texttt{L050AGN} (green) simulation sets for comparison, as well as observational constraints from \protect\cite{baldry_galaxy_2012}.
		Lines are dotted where there are fewer than 10 galaxies per bin.
		\textit{Left:} the predicted GSMF on the $(100 \, \mathrm{Mpc})^3$ DMO volume from machine learning models trained on the \texttt{L050AGN} (purple, dashed) and \texttt{L050AGN+Zoom} (purple, solid) simulation sets.
		\textit{Right:} the predicted GSMF on the $(800 \; \mathrm{Mpc})^3$ P-Millennium DMO simulation, from machine learning models trained on the \texttt{L050AGN} (blue, dashed) and \texttt{L050AGN+Zoom} (blue, solid) simulation sets.}
    \label{fig:gsmf_pmillennium}
\end{figure*}

An alternative explanation is the well known effect of baryons on their host dark matter haloes \citep[\textit{e.g.}][]{velliscig_intrinsic_2015,schaller_baryon_2015}.
This may not only affect the masses of haloes, but also their mass distribution, changing the substructure on small scales, and hence the clustering measurement \citep{van_daalen_impact_2014,hellwing_effect_2016}.
To test whether this is causing the lower normalisation at small scales, we extract a catalogue of features from the full hydro simulation (\texttt{L100Ref}) and use these as inputs to the \texttt{L050AGN+Zoom} model.
We emphasise that these 'halo` features contain the contribution from both baryons and dark matter, but are otherwise identical to the features from a DMO simulation.
The predicted clustering for this hybrid model application is shown in the second panel of \fig{clustering}; the normalisation matches that of the \texttt{L100Ref} simulation, confirming that it is indeed baryonic effects causing the lower normalisation on small scales.
We stress that this is not strictly a fair use of the machine learning model, as it was trained on haloes from a DMO simulation, and as such the predictions should be taken with some caution.
However, we argue this is a relatively `clean' test of the impact of baryons on the halo, and the knock on effect on the clustering.

Other effects may also contribute to the deficit, such as differences in the parameters of the halo finder between DMO and hydro simulations, and for different resolution simulations.
However, it seems clear that baryonic effects on haloes are a key contributor.
A similar effect at small scales has been seen in semi-analytic models applied to DMO simulations \citep{farrow_galaxy_2015,contreras_galaxydark_2015}.
The machine learning model presented here allows us to cleanly test this effect on identical haloes.

We also compared the model predictions for the projected correlation function against those using a single subhalo feature (subhalo mass) to predict the stellar mass, applied to the P-Millennium volume.
The normalisation is underestimated in this simple model compared to the GAMA measurements, and this is particularly pronounced in the highest stellar mass bin.
Full details are provided in \app{shamcomp}.

\subsection{The Galaxy Stellar Mass Function}

\begin{figure*}
	\includegraphics[width=\textwidth]{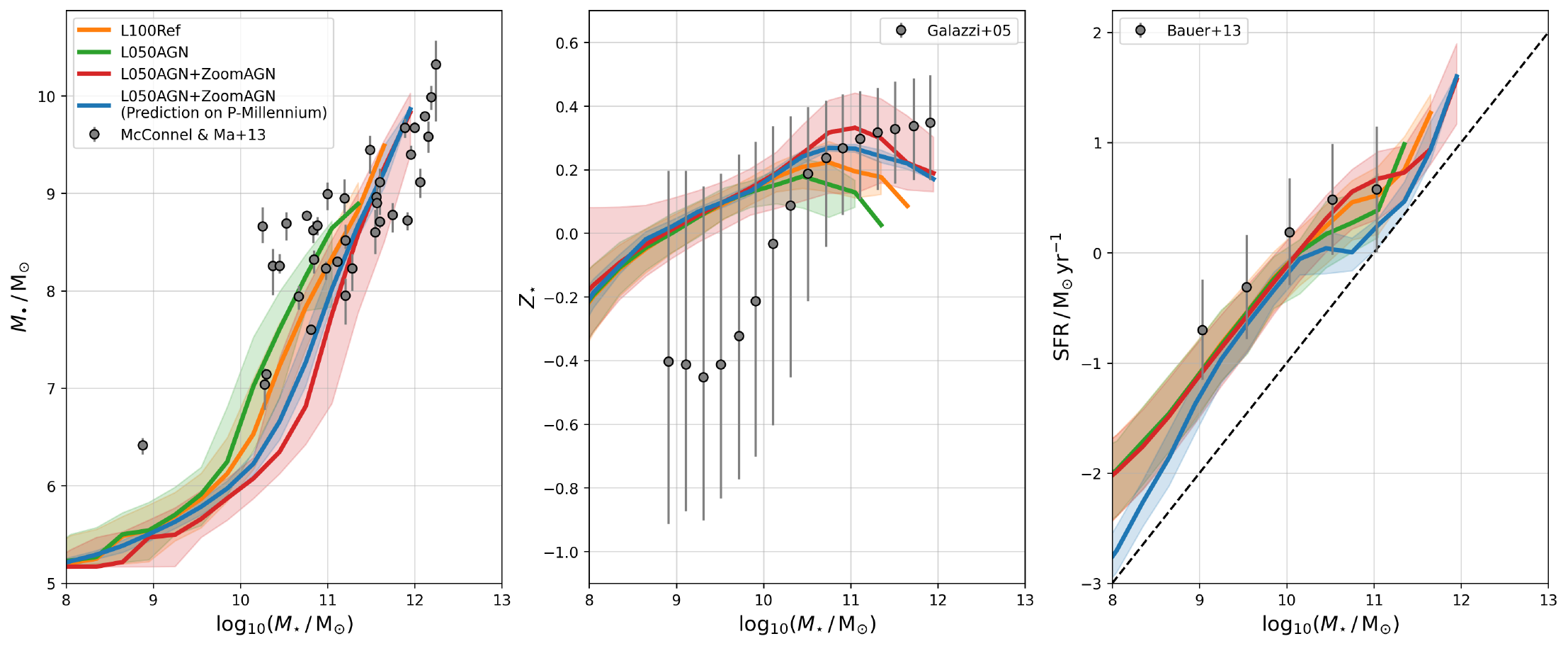}
    \caption{The black hole -- stellar mass relation (left), stellar mass -- metallicity relation (middle) and star forming sequence (right).
		Observational constraints for each relation are shown, from \protect\cite{mcconnell_revisiting_2013}, \protect\cite{gallazzi_ages_2005} and \protect\cite{bauer_galaxy_2013}, respectively.
		The relation in the \texttt{L100Ref} (orange), \texttt{L050AGN} (green) and \texttt{L050AGN+Zoom} (red) simulation sets is shown, as well as the predicted relation from the \texttt{L050AGN+Zoom} machine learning model applied to the P-Millennium simulation (blue).
		The median is given by the solid line in each case, and the $16^{\mathrm{th}}-84^{\mathrm{th}}$ percentile range is shown by the coloured shaded region.
		The dashed black line in the right panel shows the cut used for passive galaxies, $\mathrm{sSFR} < 10^{-11} \, \mathrm{yr^{-1}}$.
		}
    \label{fig:pmillennium_dfs}
\end{figure*}

The left panel of \fig{gsmf_pmillennium} shows the \texttt{L050AGN} model run on the \texttt{L100Ref} DMO simulation.
We compare to the GSMF from the hydrodynamic \texttt{L100Ref} simulation, and it is clear that the high mass end of the GSMF is not reproduced.
Whilst there are parameter differences between the models, it is not expected that the AGNdT9 model would fail to produce any $10^{12} \, \mathrm{M_{\odot}}$ galaxies in a $(100 \, \mathrm{Mpc})^3$ volume.
In fact, the predictions broadly follow the model used for training, \texttt{L050AGN}, though underestimate the abundance of galaxies at the high-mass end ($> 10^{11} \, \mathrm{M_{\odot}}$).
This additional underestimate is likely the result of a lack of training data at the high-mass end, due to the low number of high-mass galaxies in the \texttt{L050AGN} volume.

However, if we use the \texttt{L050AGN+Zoom} model we get much better agreement with the \texttt{L100Ref} simulation at the high-mass end.
This demonstrates the effect of including the \ceagle\ zoom regions; the model is able to learn the baryonic properties of galaxies in the cluster regions, which are not present in \texttt{L050AGN}.
Predictions at lower stellar masses are also consistent with both \texttt{L100Ref} and \texttt{L050AGN} down to $\sim 10^8 \; \mathrm{M_{\odot}}$, the approximate resolution limit of the original simulations \citep{schaye_eagle_2015}, and where our predictions are approximately complete (see \app{completeness}).

We now turn our attention to the much larger \pmill\ DMO simulation.
The right panel of \fig{gsmf_pmillennium} shows predictions for the \texttt{L050AGN} and \texttt{L050AGN+Zoom} models on this volume, and whilst the former still completely misses the high mass end, the model including zooms is able to predict stellar masses out to $\sim 10^{12} \, \mathrm{M_{\odot}}$.
This extends the dynamic range of the GSMF beyond that accessible to the \texttt{L100Ref} hydrodynamic simulation, and improves the statistics significantly.
This is a significant achievement of the model -- it is able to successfully extend the predictive range beyond that achievable with periodic hydrodynamic simulations.
At lower stellar masses the predictions are consistent with both \texttt{L100Ref} and \texttt{L050AGN}.
The predictions at the high mass end are also in broad agreement with the observational constraints from \cite{baldry_galaxy_2012}.

The P-Millennium simulation is lower resolution than those used for training, which may impact the predicted properties of galaxies, particularly those close to the resolution limit.
To test the impact of resolution, we applied the \texttt{L050AGN+Zoom} model to a lower resolution $(100 \mathrm{Mpc})^3$ DMO run, with 8 times fewer particles.
The predictions for the GSMF were identical, which confirms that differing resolution has no impact on the predicted properties; as long as the haloes are resolved, the halo features used for prediction are robust.

\subsection{The black hole -- stellar mass relation}
We have demonstrated how the model is able to predict stellar masses with high accuracy, and produce a GSMF for the \pmill\ simulation volume.
We now explore other key baryonic distribution functions.
\fig{pmillennium_dfs} shows the black hole -- stellar mass relation, the stellar mass -- metallicity relation, and the star-forming sequence.
Each panel shows the relation in the \texttt{L100Ref}, \texttt{L050AGN} \& \texttt{L050AGN+Zoom} simulations, as well as the predicted relation for our  \texttt{L050AGN+Zoom} model, with fiducial feature set, run on the \pmill\ simulation.

The black hole -- stellar mass relation shows a rapid increase in the stellar mass above $M_{\star} \sim 10^{10} \; \mathrm{M_{\odot}}$, though the exact mass at which the relation turns upwards is dependent on the simulation.
In \texttt{L050AGN+Zoom} the increase is at a higher mass compared to the two periodic simulations.
This is not due to any parameter differences, since \texttt{L050AGN} has identical parameters, but may be due to the cluster environment delaying black hole accretion by starving the central regions of a galaxy of gas.
Though \cite{van_son_galaxies_2019} note an excess of `black hole monster galaxies' in cluster environments due to tidal stripping, this is a sub-dominant population compared to the main relation, so it does not increase the normalisation of the black hole -- stellar mass relation in these environments.
The model predictions lie between the periodic and zoom relations, which is perhaps expected since both environments are providing training data from which the machine is making its predictions.
Overall the relation is predicted remarkably well, and the predictions extend the dynamic range to higher stellar and black hole masses than those achievable in \texttt{L100Ref} \& \texttt{L050AGN}.
At these higher masses the model is in good agreement with the observational results of \cite{mcconnell_revisiting_2013}, though the scatter at fixed stellar mass is still underpredicted \citep[as seen in][]{schaye_eagle_2015}.

\subsection{The stellar mass -- metallicity relation}
Predictions from the model for the stellar mass -- metallicity relation show similar behaviour.
The model predictions lie between the relations from the periodic and zoom simulation sets at high stellar masses ($M_{\star} \,/\, \mathrm{M_{\odot}} > 10^{10}$), but closely follow the predictions below this, except at the very lowest stellar masses.

The scatter in both these relations is much tighter for the model predictions than in the original simulation sets.
This is a reflection of the deterministic nature of the machine learning prediction, combined with the relatively limited feature set, which has been discussed in a number of previous works \citep[\textit{e.g.}][]{kamdar_machine_2016-1,moews_hybrid_2020}.
Historical halo features, such as the formation and assembly time, may help to increase the diversity of baryonic properties at fixed stellar mass.
However, the predictions still lie within the uncertainties on observational constraints from \cite{gallazzi_ages_2005} at all stellar masses.

\subsection{The star forming sequence}
Finally, the right panel of \fig{pmillennium_dfs} shows the star-forming sequence, excluding passive galaxies ($\mathrm{sSFR} < 10^{-11} \, \mathrm{yr^{-1}}$).
As shown in \fig{violins_SFR_L0050N0752} the model tends to underpredict SFRs of star forming galaxies, and this is reflected in the star-forming sequence, where the normalisation at $M_{*} \,/\, M_{\odot} = 10^{11}$ is lower than that in the original simulation sets, by up to -0.8 dex compared to \texttt{L050AGN+Zoom}.
The scatter at fixed stellar mass is comparable to the simulation sets ($\pm 0.25 \, \mathrm{dex}$), though this may be partly due to truly quiescent galaxies in the simulation sets that have residual star formation when predicted in the model.
In general, however, the star forming sequence is broadly reproduced, is in good agreement both above and below the characteristic mass, and lies within the uncertainties on observational constraints from \cite{bauer_galaxy_2013}.

\section{Feature exploration}
\label{sec:feature_exploration}

\begin{figure}
	\includegraphics[width=\columnwidth]{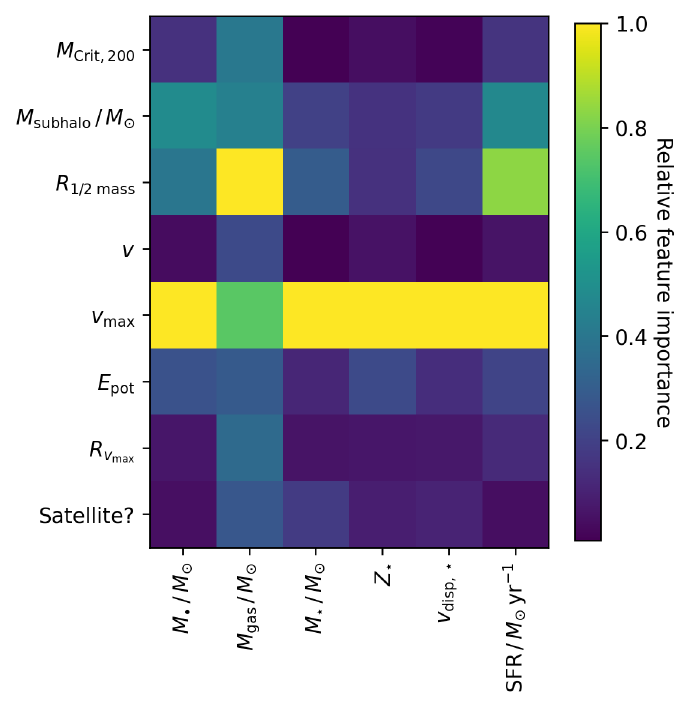}
    \caption{Matrix showing the relative importance ($0\rightarrow1$, low to high importance) of each feature ($y$-axis) for each predictor quantity ($x$-axis), for the \texttt{L050AGN+Zoom} model.
		The importance is normalised by the maximum for each predictor.
		}
    \label{fig:feature_importance_predictors}
\end{figure}

\begin{figure}
	\includegraphics[width=\columnwidth]{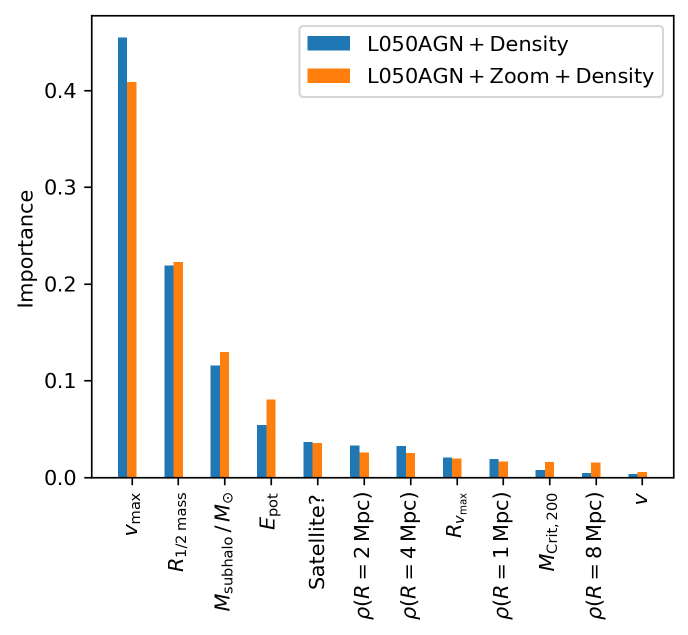}
    \caption{Relative feature importance as described by ERT, across all features simultaneously.
		\texttt{L050AGN} (blue) and \texttt{L050AGN+Zoom} (orange) machine learning models are shown, including additional features describing the local density on different scales ($\rho(R)$).
		}
    \label{fig:feature_importance_all}
\end{figure}

Feature importance in ERT can be evaluated from the relative position of a given feature in the tree; the closer to the root node in the ensemble of trees, the higher the importance.
In order to evaluate the feature importance for each predictor, we re-train the model on each predictor individually.
\fig{feature_importance_predictors} shows a matrix of each predictor against each feature, coloured by their relative importance.
The order of relative importance is generally the same for all predictors.
$V_{\mathrm{max}}$ is by far the most important feature; \cite{kamdar_machine_2016-1} attributed a similarly high importance for  $V_{\mathrm{max}}$ in their machine learning model trained on Illustris.
A number of other studies have highlighted the importance of $V_{\mathrm{max}}$ for predicting baryonic properties.
\citep{matthee_origin_2017} showed that, in \eagle, $V_{\mathrm{max}}$ is a key predictor of the stellar mass, more so than the halo mass.
\cite{chaves-montero_subhalo_2016} use a subhalo abundance matching technique to test the recovery of the clustering of galaxies in \eagle\ and find a similarly strong dependence on $V_{\mathrm{max}}$.
The circumgalactic medium mass fraction, at fixed halo mass, has also been shown to correlate strongly with $V_{\mathrm{max}}$ (when parametrised as a ratio with the virial circular velocity, closely related to the halo binding energy), in both \eagle\ and Illustris \citep{davies_gas_2018,davies_quenching_2020,oppenheimer_feedback_2019}; the authors of these studies argue that a high $V_{\mathrm{max}}$ corresponds to an early collapse time for a halo, which leads to greater black hole growth, which in turn ejects more of the circumgalactic medium mass.
This has a big impact on the latter baryonic properties of the galaxy, such as its star formation history and morphology.
This explains the strong importance of $V_{\mathrm{max}}$ in our feature set for the majority of our baryonic predictors.

Interestingly, for the gas mass, the half-mass radius is instead the most important feature.
$V_{\mathrm{max}}$ is still of high importance, but at a similar level to the subhalo mass and total halo mass ($M_{\mathrm{crit},200}$).
This suggests that the size of the underlying dark matter halo is closely related to its current gas mass.
A similarly strong correlation between (HI) gas mass and size has been found observationally, though with the stellar component rather than dark matter \citep{catinella_galex_2012}.

The peculiar velocity is the least important feature for all predictors, as expected.
Interestingly, features that encode the local halo environment, such as $M_{\mathrm{crit,200}}$ and its status as a satellite or central, are also two of the least important features.
This suggests that the properties of the subhalo itself mostly determine the baryonic properties, however this does not necessarily mean that `nature' rather than `nurture' is the dominant evolutionary process.
Instead, other subhalo features may encode environmental information, \textit{e.g.} satellites are clear outliers in the $M_{200} - M_{\mathrm{subhalo}}$ plane.

We also evaluated the effect of including local density features, $\rho(R)$.
\fig{feature_importance_all} shows the feature importance for all predictors, in the \texttt{L050AGN} and \texttt{L050AGN+Zoom} machine learning models.
None of these local features dominates the feature importance, but the density on intermediate scales ($R = [2,4] \; \mathrm{Mpc}$) has a higher importance than on the smallest and largest scales ($R = [1,8] \; \mathrm{Mpc}$, respectively).
The order of feature importance is otherwise mostly preserved.

\section{Discussion \& Conclusions}
\label{sec:conc}

We have demonstrated the effectiveness of machine learning methods in modelling the complex relationships between galaxies and their host haloes by training a machine learning model to directly learn this mapping.
By combining hydro and DMO simulations we avoid baryonic effects on haloes that would bias predictions.
And by using a training set consisting of both periodic and zoom simulations of galaxy clusters, we include rare environments that may not be present in typical periodic simulations, allowing the model to be applied to much larger volume dark matter-only simulations, increasing the dynamic range, and allowing the evaluation of clustering statistics over much larger scales.
Our conclusions are as follows:
\begin{itemize}
  \item The model successfully predicts the stellar mass, stellar velocity dispersion and black hole mass, and provides reasonable predictions for the star formation rate, stellar metallicity and total gas mass.
  Even where the stellar metallicity shows some dispersion in the prediction, the overall distribution is recovered.
  \item Star formation rates and gas masses are biased low due to the effect of quiescent, gas poor galaxies, and some suggestions for improving this are put forward, including the use of historical halo features.
  \item Adding features representing the local density leads to a negligible increase in the predictive accuracy for most properties, except the gas mass, which shows significant improvement, particularly in cluster environments.
  \item We apply the trained model to the \pmill\ simulation and analyse the projected two-point correlation function.
  We are able to predict the clustering of galaxies out to much larger scales than in the periodic hydro simulations ($> 10 \, h^{-1} \, \mathrm{Mpc}$), as well as analyse the clustering of rarer, high-mass galaxies, and find that \eagle\ is in good agreement with observational constraints from GAMA on large scales.
  On smaller scales we conclude that baryonic effects on haloes affect the clustering statistics.
  \item The predicted galaxy stellar mass function is in excellent agreement with that given by the periodic hydro simulations at low and intermediate masses, and extends the relation to higher masses.
  \item The black hole -- stellar mass and stellar mass -- metallicity relations are well reproduced, though with less scatter, as seen in other machine learning models.
  \item The normalisation of the star-forming sequence is slightly under-predicted at the characteristic mass, which reflects both the lower normalisation in the training data, but also the lower predicted stellar masses on the test set.
  However, the general form is in good agreement.
  \item $V_{\mathrm{max}}$ is the most important feature in all simulation sets. Measures of the local environment, such as the satellite flag, host halo mass and local density, do not show high importance in any of the models.
\end{itemize}

We stress that our model is not intended as a replacement of traditional galaxy formation models: it is in fact wholly reliant on such models to train from.
It does, however, provide a means of expanding the predictions from such models to much larger periodic volumes.
These larger volumes are useful for a number of science questions.
Galaxy clustering is a particularly important application we have demonstrated here, allowing us to test the clustering statistics of high-resolution hydrodynamic simulations in the high-mass, large-separation regime.
As demonstrated by \cite{jo_machine-assisted_2019}, additional features, such as the halo merger history and its assembly and formation time, are expected to have a significant positive impact on the prediction accuracy.
Whilst we have found that features describing the local environment are not highly important, additional parameters describing, for example, the tidal shear \citep[\textit{e.g.}][]{lucie-smith_interpretable_2019}, may also encode more useful information for the machine to learn from.
It may also be possible to make predictions at multiple redshifts simultaneously by providing the machine with the scale factor, as demonstrated in \cite{moster_galaxynet_2020}.

The \textsc{C-EAGLE} sample provides a wealth of training data on rich cluster environments, however those environments on the opposite end of the overdensity distribution, extreme underdensities or \textit{cosmic voids}, are less well sampled in our training set.
Void regions do not have as obvious an effect on their constituent galaxies properties as rich cluster environments, where galaxy mergers are far more common and extreme processes such as ram-pressure stripping occur, however noticeable effects are still seen in voids in the fiducial periodic \textsc{EAGLE} volumes \citep{paillas_baryon_2017,xu_galaxy_2020}.
Larger, more significantly underdense regions are, as for overdense regions, not well sampled in the periodic volumes, however such voids are an important constituent of the universe, making up $\sim$ 60\% of the cosmic volume \citep{pan_cosmic_2012}.
In future work we will use resimulations of a range of overdensities down to low redshift to better populate this region of overdensity space, reducing generalization errors for galaxies in these environments.

We have focused on six key baryonic properties, but other baryonic properties are simple to add, as well as the emission properties of galaxies if combined with post-processing pipelines.
This will allow for the construction of extremely large lightcones \citep[as demonstrated in][using their empirical modelling plus simulation-calibrated approach]{hearin_generating_2020}, necessary for making predictions for wide field surveys from the upcoming Roman and Euclid space-based observatories \citep{potter_pkdgrav3_2017}.
To this end, in future work we will explore predictions during the epoch of reionisation, where we will leverage the \flares\ simulations \citep{lovell_first_2021}.
A unique aspect of \flares\ is that it consists of resimulations of a range of overdensities, providing training data in extreme over- and \textit{under}-dense environments, which may aid predictions of galaxy properties across all environments.

\section*{Acknowledgements}
The authors wish to thank the anonymous referee for a detailed report that improved this manuscript.
We also wish to thank Daniel Farrow for providing the GAMA clustering data, John Helly for help reading the P-Millennium data, and Aswin Vijayan, Rob Crain, Joop Schaye and Scott Kay for helpful comments and discussions.
Thanks also to the \eagle\ team for their efforts in developing the \eagle\ simulation code.

We acknowledge the following open source software packages used in the analysis: \textsf{scipy} \citep{2020SciPy-NMeth}, \textsf{Astropy} \citep{robitaille_astropy:_2013} , \textsf{matplotlib} \citep{Hunter:2007}, \textsf{seaborn} \citep{waskom_seaborn_2021} and \textsf{halotools} \citep[][v0.7]{hearin_forward_2017}.

This work used the DiRAC@Durham facility managed by the Institute for Computational Cosmology on behalf of the STFC DiRAC HPC Facility (www.dirac.ac.uk).
The equipment was funded by BEIS capital funding via STFC capital grants ST/K00042X/1, ST/P002293/1, ST/R002371/1 and ST/S002502/1, Durham University and STFC operations grant ST/R000832/1.
DiRAC is part of the National e-Infrastructure.
PAT acknowledges support from the STFC (grant number ST/P000525/1).
CCL acknowledges support from the Royal Society under grant RGF/EA/181016.
CMB acknowledges support from STFC grant ST/T000244/1.
MS is supported by the Netherlands Organisation for Scientific Research (NWO) through VENI grant 639.041.749.
GF acknowledges the support of the European Research Council under the Marie Sk\l{}odowska Curie actions through the Individual Global Fellowship No.~892401 PiCOGAMBAS.
YMB gratefully acknowledges funding from the NWO through VENI grant number 639.041.751.

\section*{Data Availability}
The public \eagle\ database can be used to access the subhalo properties for the periodic hydrodynamic simulations in this paper \citep{mcalpine_eagle_2016}.
Other data underlying this article will be shared on reasonable request to the corresponding author.
The code used to train and analyse the models, and produce all plots, is made available at \href{https://github.com/christopherlovell/ML-cosmo}{github.com/christopherlovell/ML-cosmo}.




\bibliographystyle{mnras}
\bibliography{ML_sims,custom} 



\appendix

\section{Stellar mass completeness}
\label{sec:completeness}

\begin{figure}
	\includegraphics[width=\columnwidth]{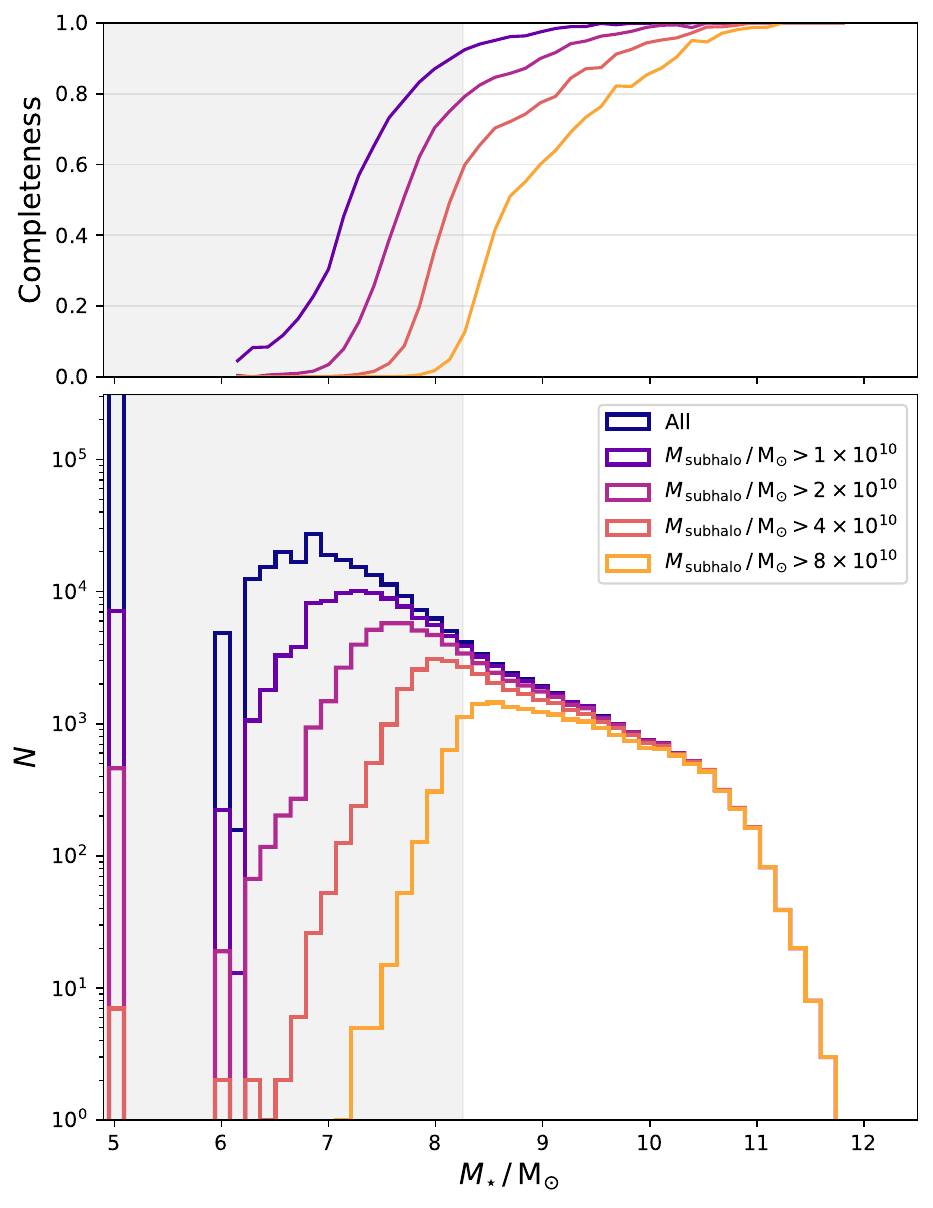}
    \caption{Histogram of stellar masses in the \texttt{L100Ref} simulation for different DMO subhalo mass limits from the matched subhalo list.
    The grey shaded area designates the stellar mass resolution limit ($M_{\star} \,/\, \mathrm{M_{\odot}} > 1.8 \times 10^{8}$, or 100 star particles at the initial baryon mass).}
    \label{fig:stellar_mass_completeness}
\end{figure}

Before model training we pre-select haloes based on their dark matter properties only, to ensure the same selection can be applied to any DMO simulation the model is applied to.
This is intended to avoid a situation where a model is applied to haloes with properties that were not present in the training set.
Since the selection is done on DMO properties only, we here check whether galaxies below the resolution limit in the hydro simulation are included, and the incompleteness of galaxies above the resolution limit.
\fig{stellar_mass_completeness} shows a histogram of stellar mass in the \texttt{L100Ref} simulation for different DMO subhalo mass cuts.
For even the strictest subhalo mass limit there are large numbers of subhaloes with stellar masses below the resolution limit; this suggests their baryonic properties are highly unresolved.
However, the important quantity is the completeness at fixed stellar mass.
For a subhalo mass limit of $M_{\mathrm{subhalo}} \,/\, \mathrm{M_{\odot}} > 10^{10}$ the completeness is greater than 95\% above the stellar mass resolution limit ($M_{\star} \,/\, \mathrm{M_{\odot}} > 1.8 \times 10^{8}$, approximately equal to 100 star particles at the initial baryon mass, i.e. ignoring stellar evolution mass loss), and 100\% complete above $5 \times 10^{9} \, \mathrm{M_{\odot}}$.
We use a subhalo mass limit of $M_{\mathrm{subhalo}} \,/\, \mathrm{M_{\odot}} > 10^{10}$ throughout the rest of the text.

\section{Isotonic Fits to a Single Feature}
\label{sec:shamcomp}

\begin{figure}
	\includegraphics[width=\columnwidth]{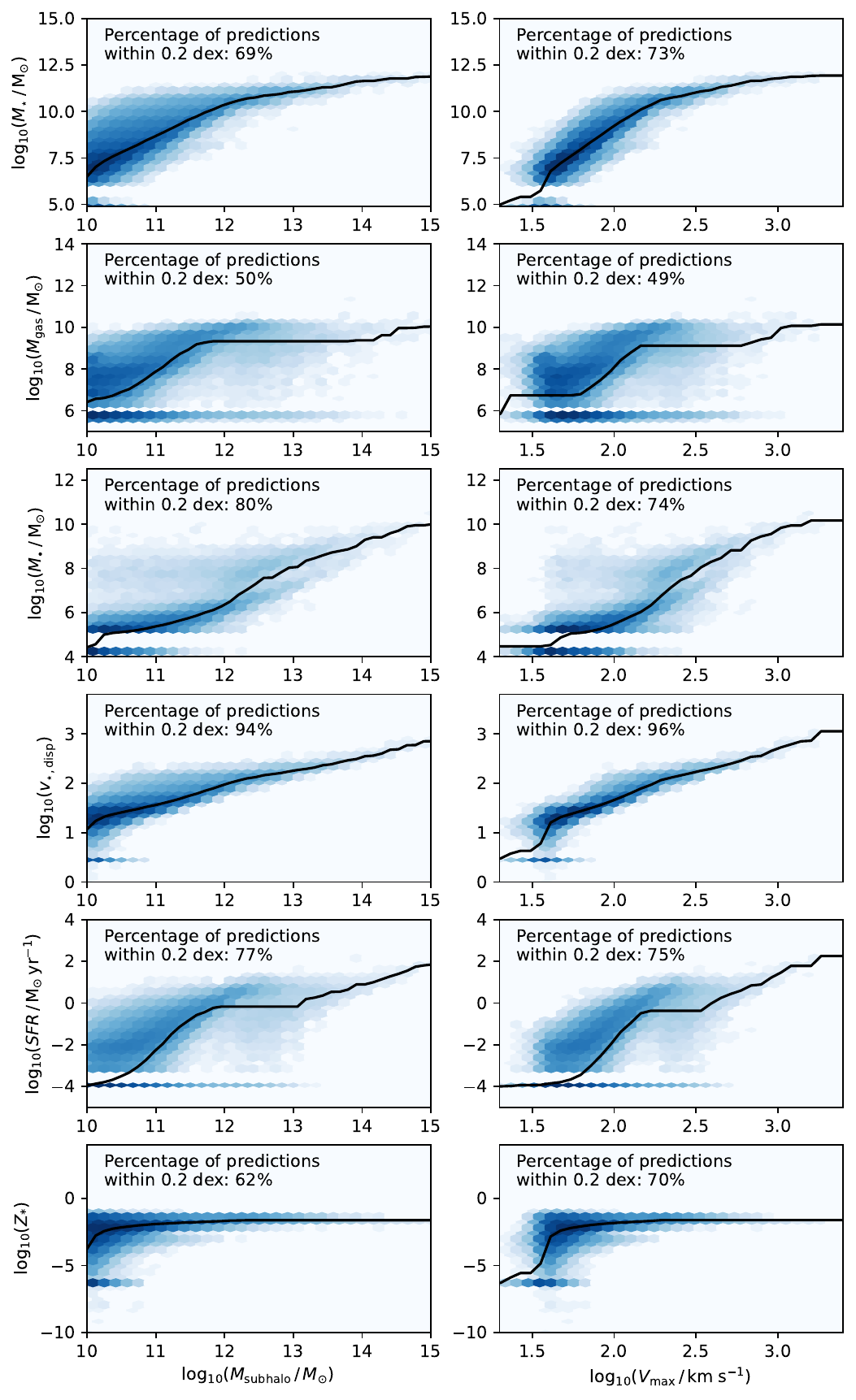}
    \caption{
		Relations between features commonly used in SHAM approaches (Subhalo mass and $V_{\mathrm{max}}$; $x$--axis) and each predictor ($y$--axis).
		Each panel shows a 2D histogram of the distribution (blue) alongside a fitted monotonic linear relation (black line).
		The percentage of galaxies where the predicted value is within 0.2 dex of the true value is quoted in each panel.
		}
    \label{fig:sham_comparison}
\end{figure}

In order to provide a qualitative assessment of the ERT model we choose to fit a simple model to the relationship between each predictor and a \textit{single} feature.
We use subhalo mass and $V_{\mathrm{max}}$ as our chosen features as these are commonly used in SHAM approaches.
We fit each relation with an Isotonic regression model, which ensures monotonicity.
We do this for the training set, and evaluate the performance on the test set.
Each relation, and the corresponding fits, are shown in \fig{sham_comparison}.
The percentage of galaxies where the predicted value is within 0.2 dex of the true value is quoted in each panel.
In each case this percentage is lower than that achieved with the ERT model.

\begin{figure}
	\includegraphics[width=\columnwidth]{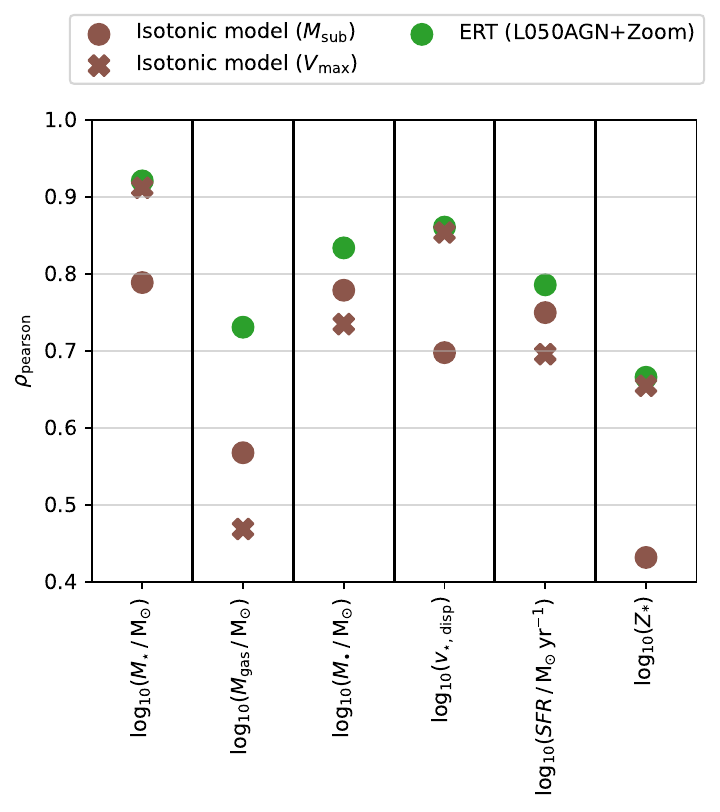}
    \caption{
		Pearson correlation coefficient for the ERT model (\texttt{L050AGN+ZoomAGN}) as well Isotonic regression models trained using subhalo mass and $V_{\mathrm{max}}$.
		Each predictor is shown on the $x$-axis.
		}
    \label{fig:pearson_appendix}
\end{figure}

We also show the pearson correlation coefficient for the ERT model as well as the Isotonic regression model for each feature in \fig{pearson_appendix}.
The ERT model outperforms the Isotonic regression model for all predictors, though the performance is comparable using $V_{\mathrm{max}}$ for the stellar mass, stellar velocity dispersion, and stellar metallicity.
This is expected from the strong correlation between the predictor and $V_{\mathrm{max}}$ in each of these cases, shown in \fig{sham_comparison}.
\fig{feature_importance_predictors} also shows that these three predictors are particularly dependent on $V_{\mathrm{max}}$, whereas other predictors have greater contributions from other features.
It is also interesting to see that subhalo mass is the more accurate predictor for gas mass, black hole mass and star formation rate compared to $V_{\mathrm{max}}$, which highlights that using one or the other feature in a SHAM approach may not lead to optimised predictions for all galaxy features -- the ML approach, on the other hand, simply incorporates all features, and chooses the best for each predictor.

\begin{figure}
	\includegraphics[width=\columnwidth]{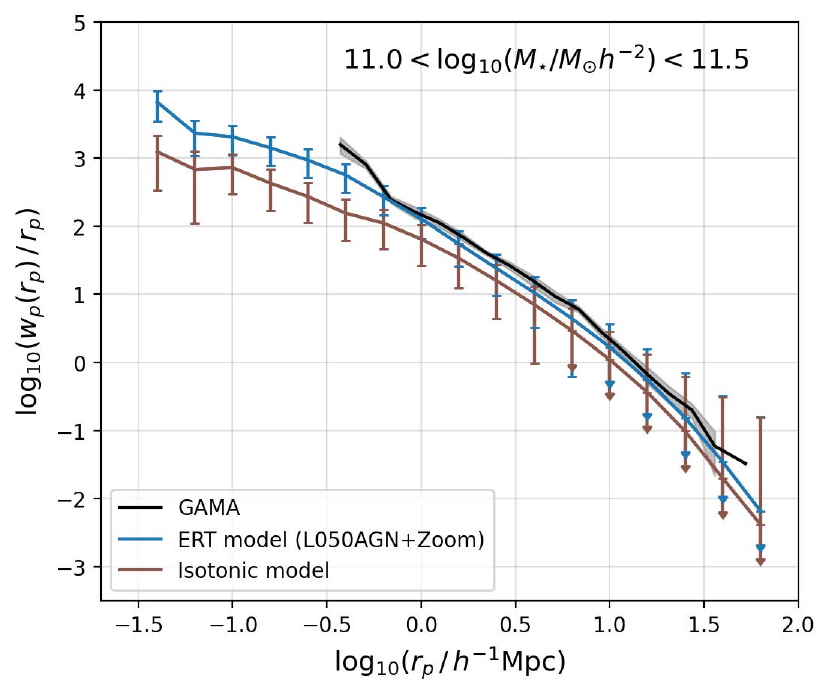}
	\includegraphics[width=\columnwidth]{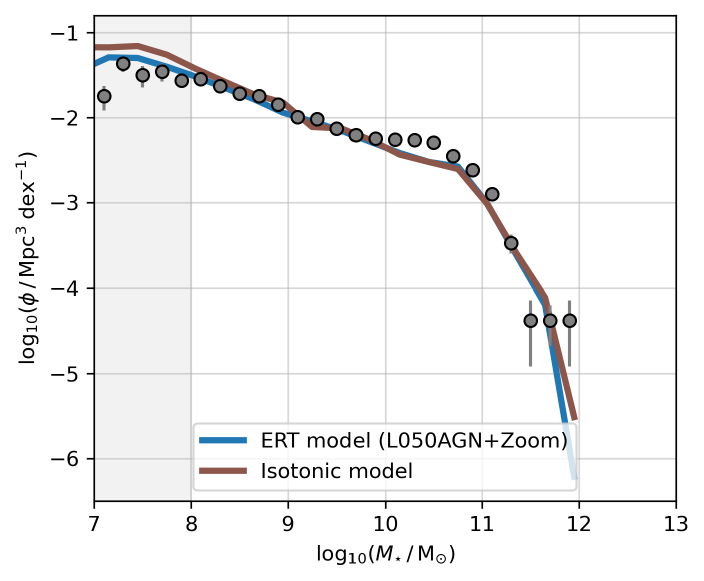}
    \caption{
		Predictions for the projected correlation function (top panel) and galaxy stellar mass function (bottom panel) using the Isotonic regression model (using subhalo mass; brown lines), compared to the ERT model (blue) with all features.
		}
    \label{fig:clustering_appendix}
\end{figure}

In \fig{clustering_appendix} we show the impact of using the Isotonic regression model (using subhalo mass as the feature) on the projected correlation function and the GSMF.
The GSMF is mostly reproduced, as expected due to the strong correlation between feature and predictor.
However, the projected correlation function (for $11 < M_{\star} \,/\, \mathrm{M_{\odot}} < 11.5$) shows a deficit in the normalisation compared to the ERT model, particularly on small scales.
One explanation is that high mass satellite galaxies, which are not common in the training set, may be more common in the larger P-Millennium volume.
The ERT model then handles these objects better than the Isotonic model, utilising other features that are more important in these environments (\textit{e.g.} the satellite flag).

\bsp	
\label{lastpage}
\end{document}